\documentclass[fleqn,usenatbib]{mnras}

\usepackage{newtxtext,newtxmath,float}
\usepackage[T1]{fontenc}

\DeclareRobustCommand{\VAN}[3]{#2}
\let\VANthebibliography\thebibliography
\def\thebibliography{\DeclareRobustCommand{\VAN}[3]{##3}\VANthebibliography}

\usepackage{graphicx}	% Including figure files
\usepackage{amsmath}	% Advanced maths 
\usepackage{bm}
\usepackage[dvipsnames]{xcolor}
\usepackage{orcidlink}
\usepackage{balance}

\hypersetup{
    colorlinks = true,
    citecolor = {MidnightBlue},
    linkcolor = {BrickRed},
    urlcolor = {BrickRed}
}

\newcommand{\gvec}{\mathbfit{g}}
\newcommand{\nvec}{\mathbfit{n}}

\newcommand{\svec}{\mathbfit{s}}
\newcommand{\dvec}{\mathbfit{d}}
\newcommand{\mvec}{\mathbfit{m}}
\newcommand{\rvec}{\mathbfit{r}}

\newcommand{\Xmtrx}{\mathbfss{X}}

\newcommand{\Pmtrx}{\mathbfss{P}}
\newcommand{\Bmtrx}{\mathbfss{B}}
\newcommand{\Gmtrx}{\mathbfss{G}}
\newcommand{\Nmtrx}{\mathbfss{N}}
\newcommand{\Mmtrx}{\mathbfss{M}}
\newcommand{\Kmtrx}{\mathbfss{K}}
\newcommand{\Smtrx}{\mathbfss{S}}

\newcommand{\ghat}{\hat{\bm{g}}}

\title[Bayesian recalibration of flux scale factors in diffuse radio maps]{Bayesian recalibration of flux scale factors in diffuse radio maps using low-resolution absolute radiometers}

\author[A. Nasirudin \& P. Bull]{
Ainulnabilah Nasirudin$^{1}$\thanks{E-mail: ainulnabilah.nasirudin@manchester.ac.uk} \orcidlink{0000-0003-2213-4547}
and Philip Bull\,\orcidlink{0000-0001-5668-3101}$^{1,2}$
\\
$^{1}$Jodrell Bank Centre for Astrophysics, University of Manchester, Manchester, M13 9PL, United Kingdom\\
$^{2}$Department of Physics and Astronomy, University of Western Cape, Cape Town 7535, South Africa\\
}

\date{Accepted XXX. Received YYY; in original form ZZZ}

\pubyear{\the\year{}}

\begin{document}
\label{firstpage}
\pagerange{\pageref{firstpage}--\pageref{lastpage}}
\maketitle

% Abstract of the paper
\begin{abstract}
The Haslam 408 MHz all-sky map is widely used as a template to model the diffuse Galactic synchrotron emission at radio and microwave frequencies. Recent studies have suggested that there are large uncorrected flux scale errors in this map, however. We investigate the possibility of statistically recalibrating the Haslam map using absolutely-calibrated (but low angular resolution) radio experiments designed to measure the 21cm global signal at low frequencies. We construct a Gibbs sampling scheme to recover the full joint posterior distribution of $\sim 50,000$ parameters, representing the true sky brightness temperature field, as-yet uncorrected flux scale factors, and synchrotron power-law spectral indices. With the idealised full-sky simulated data, we perform a joint analysis of a $1^\circ$ resolution diffuse map at 408~MHz and multi-band 21cm global signal data with $30^\circ$ resolution under different assumptions about 1) noise levels in the maps, 2) sky coverage, and 3) synchrotron spectral index information. For our fiducial scenario in which the global signal experiment has a 50~mK noise rms per coarse pixel in each of 20 frequency bins between 50 -- 150~MHz -- the typical range for a global signal experiment, we find that the notional Haslam flux scale factors can be recovered in most (but not all) sub-regions of the sky to an accuracy of $\pm 2 \%$. In all cases we are able to rectify the sky map to within $\sim 5$~K of the true brightness temperature. Our method can be used to correct the Haslam map once maps obtained from global experiments are available.

\end{abstract}

% Include between one and six keywords.

%%%%%%%%%%%%%%%%%%%%%%%%%%%%%%%%%%%%%%%%%%%%%%%%%%

%%%%%%%%%%%%%%%%% BODY OF PAPER %%%%%%%%%%%%%%%%%%

\section{Introduction}

The 408 MHz map from \cite{haslam1982A&AS...47....1H} comprises data from three radio telescopes -- the Jodrell Bank Mk-I 76~m (later upgraded to the Mk-IA), the Effelsberg 100~m, and the Parkes 64~m -- that have been combined into a comprehensive all-sky map of the Galactic diffuse radio emission. Since its publication more than 40 years ago, the map has been used extensively in a wide range of studies. One particularly important use is as a synchrotron template for foreground removal in cosmic microwave background (CMB) observations \citep[e.g.][]{Bennett_2003,Bennett_2013,2016A&A...594A..10P, 2025arXiv250606274B}, and in neutral hydrogen (H$_\textsc{I}$) 21cm brightness temperature surveys \citep[e.g.][]{battye_10.1093/mnras/stt1082, wolz_10.1093/mnras/stu792, 10.1093/mnras/stab1560, 2025MNRAS.537.3632M, 2025arXiv250403554H}. In the latter surveys, diffuse sky models based on the Haslam map are sometimes also used for calibration.

To build such sky models, the all-sky 408 MHz map is combined with data from other surveys at different frequencies, with a power-law frequency spectrum or similar assumed in order to scale the diffuse emission brightness temperature to different frequencies. Collections of many sky maps (most with only partial sky coverage) across a wide range of frequencies have been used to construct `global' radio sky models with complex frequency spectra corresponding to multiple physical emission components. These include the Global Sky Model \citep[GSM; ][]{gsm10.1111/j.1365-2966.2008.13376.x}, the improved GSM \citep{gsmimproved10.1093/mnras/stw2525}, the Global Model for the Radio Sky Spectrum \citep[GMOSS; ][]{gmoss2017AJ....153...26S}, the Python Sky Model \citep{2025arXiv250220452T}, and CosmoGlobe \citep{2023A&A...679A.143W}. In all of the above, the Haslam map plays an important role, either in determining the spatial distribution of the synchrotron emission over the full sky, anchoring the frequency spectrum, or both.

Despite the availability of other all-sky models derived from more recent radio sky surveys, e.g. the Low Frequency Sky Model (LFSM) \citep{lfsm10.1093/mnras/stx1136} -- see also \citet{2021MNRAS.505.1575S} -- the Haslam map continues to be used as the main template in both observation and theoretical studies because it is the only high-resolution, full-sky synchrotron tracer map available. However, it is widely known that there are artefacts in the map from imperfectly subtracted extragalactic sources and large-scale stripes attributable to the scanning pattern of the original surveys. These have been corrected to a large extent in  \cite{remazeilles_10.1093/mnras/stv1274}. The use of the Haslam map in multi-frequency microwave experiments such as Planck \citep{2016A&A...594A..10P} has also revealed an inconsistency that renders the use of a single spectral index across all frequencies inadequate without amplitude corrections to the 408 MHz map, suggesting that there may be significant curvature in the spectral index -- at least if the Haslam map is taken at face value \citep[e.g.][]{2012ApJ...753..110K, 2022MNRAS.509.4923I}; see \citet{wilensky2024bayesian} for a discussion of how systematic effects may affect the inferred spectral curvature, however.

There are expected to be uncorrected spatially-varying flux scale factors in the map that modulate the reported brightness temperature field in a poorly understood manner. A widely quoted value of the systematic uncertainty associated with this effect is 10\%, with a zero-level offset of $\pm$ 4 K \citep{haslam1982A&AS...47....1H}. While \cite{remazeilles_10.1093/mnras/stv1274} believe that the values are around $\sim 5 \%$, \cite{Monsalve_2021} suggest that they are likely to be higher, although a definitive answer was not provided because of strong degeneracies between the frequencies studied. The true systematic modulation of the Haslam map thus remains unknown. Finally, the random noise level of the Haslam map is also poorly known; simple estimates that are constant over the sky are often used, e.g. see \cite{remazeilles_10.1093/mnras/stv1274}.

Recent studies have attempted to account for the flux scale and offset factors in diffuse sky maps by either performing or simulating joint fits with recent data from low-frequency 21cm `global signal' experiments \citep{Monsalve_2021, pagano_10.1093/mnras/stad3392, carter2025bayesian, 2025arXiv250621258I} or radio intensity maps \citep{wilensky2024bayesian}. The former experiments are absolutely-calibrated radiometers that perform spectroscopic measurements over a wide band around $\sim 100$~MHz, referenced to an internal absolute temperature calibrator. Their primary purpose is to measure the sky-averaged brightness temperature of the redshifted 21cm line, expected to be of order $\sim 100$~mK in this band, including multiple absorption features that correspond to particular physical processes associated with the earliest generations of star and galaxy formation \citep{Pritchard:2010pa}. This kind of experiment has several advantages as a reference dataset: they provide a well-calibrated temperature reference, along with internally-consistent spectral information to help constrain spectral indices etc. The main disadvantages are that they cover significantly lower frequencies than the Haslam map (in a range where ionospheric effects are becoming important for example), and have very low angular resolution, of order tens of degrees. Alternatively, intensity mapping surveys with higher angular resolution but no internal absolute temperature reference can be used. These have very different systematic effects, such as striping due to correlated noise, and can be at higher frequencies (targeting late-time cosmology) or low frequency (targeting the Epoch of Reionisation and Cosmic Dawn).

%Several studies have already used 21cm global signal experiment data as part of the model selection process via the calculation of Bayesian evidence for a wide range of plausible scenarios.

An important question is how best to infer the correction factors given that the reference experiments themselves are not perfect; they have very different angular resolution to the target maps; and the `true' frequency spectrum is not known a priori, and varies spatially. To this end, several Bayesian methods have been proposed to provide robust ways of inferring the correction factors given the incomplete information available.

In \cite{pagano_10.1093/mnras/stad3392}, a Bayesian model fitting and model selection process for foregrounds in 21cm global signal experiments was carried out on a (simulated) fiducial foreground map. The map was sub-divided into a variable number of regions with their own amplitude and spectral index parameters, and as many as 32 amplitude regions and 18 spectral index regions were considered (the two types of regions do not need to coincide) -- although smaller numbers of regions were favoured by a Bayesian evidence comparison for their fiducial scenario. This allowed errors in the foreground map to be marginalised, with the aim of producing an unbiased foreground model to be removed from the 21cm global signal data. The resulting constraints on the foreground error parameters could be used to estimate a zero-level offset and spatially-varying scale factors in the foreground map, although this was not the main goal of the method.

\cite{wilensky2024bayesian} also used a Bayesian model comparison approach to explore the presence of uncorrected flux scale factors in the Haslam map when jointly analysed with data from the OVRO-LWA 73~MHz \citep{2018AJ....156...32E} and MeerKLASS \citep{2022MNRAS.509.4923I} surveys, which are not absolutely calibrated using an internal reference. They found estimates of uncorrected flux scale factors as large as 60\% in the Haslam map for the three close-together regions they studied, and compared scenarios with different levels of bias in all of the input surveys, as well as synchrotron components with and without curvature. While too localised to estimate overall corrections to the Haslam map, wider intensity mapping surveys would permit improved modelling -- as long as their own calibrations are sufficiently accurate.

\cite{carter2025bayesian} also used Bayesian model comparison, in this case to choose between different numbers of components for the foreground frequency spectra. They included spatially-constant but spectrally-varying multiplicative flux scale factor and offset correction factors for a set of multi-frequency maps that were used to constrain the sky model. The temperature maps themselves were analytically marginalised to avoid having to evaluate a very high-dimensional posterior distribution that includes parameters for the temperature values in each pixel. For the favoured model in their simulated scenario, the posterior distribution for 22 parameters was estimated, having been analytically marginalised over the map parameters for 9 maps each with a \textsc{Healpix} \textsc{nside} of 32 (i.e. 12,288 pixels per map).

\cite{2025arXiv250621258I} used a map-making approach based on pseudo-inversion of a spherical harmonic sky model out to low-$\ell$ that includes beam effects, as well as a stochastic foreground model based on random realisations of a base sky model obtained by adding independent Gaussian random noise to a pixel map of spectral indices. The main effect of the stochastic foreground model is to relieve the specific model dependence of the beam correction factor that is applied to the 21cm global signal data. This method uses tools such as maximum likelihood estimates for the spherical harmonics, and the Bayesian Information Criterion (BIC) to select between spectral models of different complexity. It also incorporates information from multiple (simulated) global signal experiments around the world.

In this paper, we also develop a Bayesian statistical method to constrain the flux scale factors of a target map -- in this case a simulated version of the Haslam map -- using data from low-frequency absolutely-calibrated radiometers as a reference. We do not do this within a model comparison framework, however. Instead, we perform a parameter estimation study using the methods of Gibbs sampling and Gaussian constrained realisations. This allows us to recover the full joint posterior distribution of a model that includes spatially varying flux scale factors, the temperature of the `true' target sky map in each pixel, and a set of spectral parameters (in this case, spatially-varying power-law spectral indices). This is an extremely high dimensional parameter space -- almost 50,000 parameters in the simulated scenario we showcase here. No analytic marginalisation is required, and so we are able to directly inspect and analyse the marginal distributions and a variety of summary statistics involving arbitrary combinations of the parameters. The statistical sampling is done in a computationally tractable manner; the `standard' scenario analysis presented below returned 1,000 samples in $\sim 24$ hours on a MacBook Pro M1 laptop for example. Our code is available as an open source package.\footnote{\url{https://github.com/BellaNasirudin/bayesian_skymap}}

Because we are using simulated data, we have not produced a corrected Haslam map with our current framework, although this can be easily done once the low-frequency maps are available. For clarity, to correct for residual systematic effects, it would be most advantageous to use the original time-ordered data, allowing effects as a function of elevation/azimuth and time of observation to be removed as well. Unfortunately, we believe that the time-ordered data are no longer available (C. Dickinson, priv. comm.), meaning that only a map-level re-analysis is possible.

The paper is organised as follows. In Section \ref{sec:methods}, we define our data model and derive the maximum a posteriori (MAP) solution and Gaussian constrained realisation (GCR) equations that are the basis of our Gibbs sampling scheme. Next, we describe the simulations used to produce our mock datasets and the priors in Section \ref{sec:simulation}. We then present the results under different assumptions about noise level, sky coverage, and spectral index information in Section \ref{sec:results} before summarising our findings in Section \ref{sec:conclusion}.

\section{Methods} 
\label{sec:methods}
In this section, we describe the mathematical formalism used to statistically sample the spatially-varying flux scale factor, corrected sky map, and synchrotron spectral index for a given target experiment. We start by defining the data model in Section \ref{sec:data} and deriving the maximum a posteriori solutions in Section \ref{sec:map_soln}. Next, we detail the steps we take in the Gibbs sampling of the joint posterior distribution of the parameters in Section \ref{sec:gibbs}. We then explain how we generate our samples using Gaussian constrained realisations in Section \ref{sec:gcr}, and include an optional step to constrain the sky spectral indices in Section \ref{sec:mcmc_beta}. For clarity, we denote vectors as lower-case letters and matrices as upper-case letters, both of which are in bold.

\subsection{Data model and posterior}
\label{sec:data}
We model the Haslam data as a vector of observed pixel values,
\begin{equation}
    \dvec_{\rm H} = (\Pmtrx\, \gvec) \circ \svec_{\rm cal} + \mathbfit{n}, \label{eq:dH}
\end{equation}
where $\svec_{\rm cal}$ is the calibrated `true' sky temperature at 408 MHz, $\gvec= 1 + \delta\gvec$ is a vector of flux scale factor amplitudes, and $\Pmtrx$ is a projection operator that defines the spatial basis of the flux scale factors (e.g. spherical harmonics or pixels/zones). In our case, $\Pmtrx$ is a Boolean matrix with size $N_{\rm pix}$ by $N_{\rm zone}$, where $N_{\rm pix}$ and $N_{\rm zone}$ are the number of pixels and flux scale factor zones respectively, and as such, $\gvec$ has length $N_{\rm zone}$. The symbol $\circ$ is used to denote an element-wise multiplication of pixel vectors. We assume a Gaussian noise contribution $\nvec$ with zero mean and covariance $\Nmtrx=\left<\nvec^{\rm T}\nvec\right>$. Note that we have not introduced an angular beam here, as we only wish to consider the calibrated sky map at the native resolution of the target map of around $1^\circ$ ($56'$ for the Haslam map).

A set of well-calibrated multi-frequency data with lower angular resolution is modelled as  
\begin{equation}
\dvec_{x,\nu} =  \mathbfss{R}_\nu \Bmtrx_\nu \svec_{\nu}  + \mathbfit{n}_\nu,
\end{equation}
where $\Bmtrx_\nu$ is an angular beam convolution operator for frequency channel $\nu$ and $\mathbfss{R}_\nu$ degrades a given map to a lower (Healpix) resolution, chosen to match the pixel resolution of the data at each frequency. The true sky map is scaled to a given frequency from the reference frequency $\nu_0 = 408$~MHz. For a power-law spectrum, this can be written as
\begin{equation}
\label{eqn:spectra}
 \svec_{\nu} = \svec_{\rm cal} \left(\frac{\nu}{\nu_0}\right)^{\beta}
\end{equation}
for a spectral index $\beta$. We will later make the spectral index a spatially-dependent quantity, so that different groups of pixels can have different (shared) spectral indices. The $\mathbfss{R}_\nu$ operator makes it possible to include data vectors $\dvec_{\rm H}$ and $\dvec_{x,\nu}$ that are not defined on the same pixel grid as one another. Note that we have assumed a flux scale factor of exactly 1 for the multi-frequency data (i.e. perfect calibration), which is a reasonable approximation if dealing with a global signal experiment that is operating close to its required calibration precision (typically, accurate to a few tens of mK or better). This assumption can be relaxed, with flux scale factors introduced for each channel of the multi-frequency data, but we omit this possibility here for the sake of simplicity.

Combining both $\dvec_{\rm H}$ and $\dvec_x$ into a single block vector of observed sky maps, each map in the joint data vector $\dvec$ can be written as
\begin{equation}
\label{eqn:data}
    \dvec_{j}= \Mmtrx_j\, \svec_{\rm cal} \left(\frac{\nu_j}{\nu_0}\right)^{\beta} +\mathbfit{n}_j,
\end{equation}
where $\Mmtrx_j = (\Pmtrx\, \gvec_j) \circ  \mathbfss{R}_j \Bmtrx_j $ is a matrix operator that degrades the resolution of a sky map vector, convolves it with beam $\Bmtrx_j$, and then multiplies the result elementwise (pixelwise) by a spatially-varying flux scale factor field $ \Pmtrx\, \gvec_j$. When $j=0$, corresponding to the target (Haslam) map, we set $\gvec_j=\gvec$, and $\mathbfss{R}_j= \Bmtrx_j=\mathbfss{I}$. When $j$ has any other value, we set $\Pmtrx\, \gvec_j=\mathbfit{1}$, $\mathbfss{R}_j= \mathbfss{R}$, and $\Bmtrx_j=\Bmtrx_{\nu}$ for $\nu = \nu_j$. For simplicity, hereafter we drop the subscript from $\svec_{\rm cal}$ and refer to the calibrated or `true' sky map simply as $\svec$.

\subsection{Maximum a posteriori solution for the flux scale factors and the sky signal}
\label{sec:map_soln}

As a first step, we focus solely on $\dvec_{\rm H}$ to estimate the flux scale factors. As such, the term $\left({\nu}/{\nu_0}\right)^{\beta}$ can be omitted. Following Bayes' theorem, the conditional distribution for $\gvec$ can be written as
\begin{equation}
    p(\gvec|\dvec_{\rm H}, \svec, \Gmtrx, \Nmtrx_0 ) \propto p(\dvec_{\rm H}|\gvec, \svec, \Gmtrx, \Nmtrx_0) p(\gvec | \Gmtrx),
\end{equation}
where $\Gmtrx$ is the prior covariance of the flux scale factor parameters. The subscript $0$ corresponds to $j=0$, i.e. the target map. The expression above does not depend on $\dvec_x$ (the observed low-resolution maps) explicitly; instead, the relevant information is contained within the current estimate of $\svec$, for which a conditional distribution is defined below.

Next, we substitute in the Haslam data model (Eq.~\ref{eq:dH}), but we define $\Kmtrx \equiv \svec\ \circ \Pmtrx$ by commuting $\gvec$ with $\Pmtrx$ and switching the order of $\Pmtrx$ and $\svec$ so that the $\gvec$ term is explicit. We also subtract off the constant part of the flux scale factor to define a `residual' data vector $\rvec_{\rm H} \equiv \dvec_{\rm H} - \Kmtrx \mathbfit{1}$. Under the assumption of a Gaussian likelihood for the data, we can then write the conditional distribution for the flux scale factor fluctuation as
\begin{equation}
\begin{split}
     p(\delta\gvec|\dvec_{\rm H}, \svec, \Gmtrx, \Nmtrx_0) & \propto \exp \left ( -\frac{1}{2} (\rvec_{\rm H}- \Kmtrx\, \delta\gvec )^{\rm T} \Nmtrx_0^{-1}(\rvec_{\rm H}- \Kmtrx\, \delta\gvec) \right ) \\
     & \times \exp \left (-\frac{1}{2} \delta\gvec^{\rm T} \Gmtrx^{-1}\delta\gvec \right ),
\end{split}
\end{equation}
where the second term in the product is a Gaussian prior on the flux scale factor parameters, assumed to have a prior mean of zero.
To find the maximum a posteriori (MAP) solution for these parameters, we calculate the first derivative of the logarithm with respect to $\delta\gvec$ and set this equal to 0, resulting in an estimate of $\delta\ghat$,
\begin{equation}
\begin{split}
    \label{eqn:map}
    \left . \frac{\partial}{\partial\delta\gvec}\right |_{\delta\gvec=\delta\ghat}  (\rvec_{\rm H}- \Kmtrx\, \delta\gvec )^{\rm T}\Nmtrx_0^{-1}(\rvec_{\rm H}- \Kmtrx\, \delta\gvec )+\delta\gvec^{\rm T}\Gmtrx^{-1}\delta\gvec &= 0 \\
    \implies - \rvec_{\rm H}^{\rm T} \Nmtrx_0^{-1}   \Kmtrx   +  \Kmtrx^{\rm T} \delta\ghat^{\rm T} \Nmtrx_0^{-1} \Kmtrx   +\delta\ghat^{\rm T}\Gmtrx^{-1} &= 0.
\end{split}
\end{equation}
We can then rearrange the terms to obtain
\begin{equation}
% \begin{split}
\label{eqn:weiner_g}
    \left (\Kmtrx^{\rm T} \Nmtrx_0^{-1} \Kmtrx  + \Gmtrx^{-1}\right ) \delta\ghat = \Kmtrx^{\rm T} \Nmtrx_0^{-1}\rvec_{\rm H}.\\
    % \ghat &= [\Mmtrx^{\rm T} \Nmtrx^{-1} \Mmtrx + \Gmtrx^{-1}]^{-1} [\Mmtrx^{\rm T} \Nmtrx^{-1}\dvec_{\rm H} + \Gmtrx^{-1} \gvec_0]\\  \end{split}
\end{equation}
Here $\delta\ghat$ is the Wiener filter solution, which we can identify with the mean of the conditional distribution. The (inverse) covariance matrix of the distribution is then $\mathbf{\Sigma}^{-1} = \Kmtrx^{\rm T} \Nmtrx_0^{-1} \Kmtrx + \Gmtrx^{-1}$. %Hereafter, we consider a broad prior on the flux scale factor parameters, such that $\Gmtrx^{-1} \to 0$, so we can neglect the corresponding terms in the equations above.

We can perform an analogous derivation for the MAP solution for the conditional distribution of the sky signal, $p(\svec|\Smtrx, \Nmtrx, \gvec, \dvec$) which, again under the assumption of Gaussian likelihoods for the data, is given by
\begin{align}
    p(\svec|\Smtrx, \Nmtrx, \gvec, \dvec) \propto & \exp \left ( -\frac{1}{2} (\dvec-\Xmtrx \svec)^{\rm T} {\mathbf N}^{-1}(\dvec-\Xmtrx\svec )\right ) \nonumber \\
    & \times \exp \left (-\frac{1}{2} \svec^{\rm T} \Smtrx^{-1}\svec \right ), \label{eqn:condprob_s}
\end{align}
where $\Smtrx$ is the prior covariance matrix of the signal vector, and we have constructed a block projection matrix with (block) elements
\begin{equation}
\Xmtrx_{ij} = \delta_{ij} \left(\frac{\nu_j}{\nu_0}\right)^{\beta} \Mmtrx_j,
\end{equation}
where $\delta_{ij}$ is the Kronecker delta function, and we recall that $\Mmtrx_j$ is a function of $\gvec$.
Repeating the same steps to find the MAP solution, we take the derivative of the logarithm of Eq.~\ref{eqn:condprob_s} with respect to $\svec$,  set the expression to 0, and rearrange the terms to find the MAP estimate $\hat{\svec}$ by solving
\begin{equation}
   \left (\Xmtrx^{\rm T} \Nmtrx^{-1}\Xmtrx + \Smtrx^{-1}\right ) \hat{\svec} = \Xmtrx^{\rm T}\Nmtrx^{-1}\dvec.
\end{equation}

\subsection{Gibbs sampling}
\label{sec:gibbs}
We would now like to estimate the joint posterior distribution, $p(\gvec, \svec, \beta \,|\, \dvec, \Nmtrx, \Smtrx$) using the Gibbs sampling algorithm, a method that iteratively samples from the conditional distribution of each subset of parameters, thereby effectively sampling from the joint posterior \citep{geman4767596, gelman1995bayesian}. In our case, the three conditional distributions sampled (represented by $\longleftarrow$) for each Gibbs iteration $i$ are
\begin{equation}
\begin{split}
    \gvec_{i + 1} &\longleftarrow p(\gvec\,|\,\svec_{i}, \boldsymbol{\beta}_{i}, \dvec_{\rm H}, \Gmtrx, \Nmtrx) \\
    \svec_{i + 1} &\longleftarrow p(\svec\,|\,\gvec_{i+1}, \boldsymbol{\beta}_{i}, \dvec, \Smtrx, \Nmtrx) \label{eq:gibbsiter}\\
    \beta_{i + 1} &\longleftarrow p(\boldsymbol{\beta}\,|\,\svec_{i+1}, \gvec_{i+1}, \dvec, \Nmtrx).
\end{split}
\end{equation}
In the first part of our paper, we have set $\beta$ to a fixed value, hence our sampling steps are limited to the two explained in Section \ref{sec:map_soln}. We note that one could also sample $\Smtrx$ as part of the Gibbs sampling process, e.g. to include estimation of the angular power spectrum, but we have left that step for future work.

\subsection{Gaussian constrained realisations}
\label{sec:gcr}

In order to generate samples from the first two conditional distributions of Eq.~\ref{eq:gibbsiter}, we use the Gaussian constrained realisation (GCR) method, whereby random unit Gaussian realisations $\boldsymbol{\omega}$ are scaled by the covariance of the Wiener filter solution and added to its mean. We can then trace the full conditional distribution by repeatedly drawing different realisations of the $\boldsymbol{\omega}$ terms and solving this equation.

%Because we have considered a broad prior of $\Gmtrx^{-1}$ for $p(\gvec|\dvec_{\rm H}, \svec, \Gmtrx, \Nmtrx_0)$, only the random realisation of the noise fluctuation term ($\boldsymbol{\omega}_n$) is kept.
For $p(\gvec|\dvec_{\rm H}, \svec, \Gmtrx, \Nmtrx_0)$, random fluctuation terms for the noise and flux scale factors are added by drawing unit Gaussian random vectors ($\omega_n, \omega_{g}$) with the appropriate dimensionality, and then scaling them by the respective covariance and added to the MAP solution, yielding
\begin{equation}
   \left (\Kmtrx^{\rm T} \Nmtrx_0^{-1} \Kmtrx + \Gmtrx^{-1} \right ) \gvec = \Kmtrx^{\rm T} \Nmtrx_0^{-1}\dvec_{\rm H} + \Kmtrx^{\rm T}  \Nmtrx_0^{-1/2} \boldsymbol{\omega}_n + \Gmtrx^{-1/2}\boldsymbol{\omega}_g.
\end{equation}
Similarly with $p(\svec|\Smtrx, \Nmtrx, \gvec, \dvec)$, we generate samples by drawing unit Gaussian random realisations for both the noise and signal fluctuation terms, ($\omega_n, \omega_{s}$), and scaling them by their respective covariances,
\begin{equation}
   \left (\Xmtrx^{\rm T} \Nmtrx^{-1}\Xmtrx + \Smtrx^{-1}\right ) {\svec} = \Xmtrx^{\rm T}\Nmtrx^{-1}\dvec + \Xmtrx^{\rm T} \Nmtrx^{-1/2}\omega_n + \Smtrx^{-1/2}\omega_{s}. \label{eq:gcr_s}
\end{equation}
Note that $\omega_n$ are different random draws for each conditional distribution.
These equations can be solved using standard linear solvers, such as the conjugate gradient method, to yield samples of $\gvec$ and $\svec$.

\subsection{MCMC sampling of the spectral index}

\label{sec:mcmc_beta}
Another important, imperfectly-known parameter to include in the inference is the spectral index that describes the frequency dependence of the emission. This parameter is non-linear in the likelihood, and so a similar GCR step to the ones above cannot be defined. Instead, we have added another (optional) step to draw samples of $\beta$ using the more general Markov Chain Monte Carlo (MCMC) method. MCMC entails sampling the distribution of a parameter based on the ansatz that its probability at iteration $i + 1$ is only dependent on the probability at $i$ \citep{metropolis1953equation, hastings1970monte}. Common algorithms, such as Metropolis-Hastings, permit general probability distributions to be explored, but scale poorly with the dimensionality of the parameter space. As such, it would not be practical to jointly sample a spectral index parameter for every pixel for instance. Instead, we assume that the sky has been divided up into a relatively small number of regions, each with its own spectral index.

Depending on the parametrisation of the flux scale factor field, and effects due to instrumental beams etc., changing the spectral index in one region could in principle affect the data model in pixels beyond that region. As such, we expect some degree of correlation between the spectral indices for different regions. This makes it important to jointly sample the conditional distribution for the set of $\beta$ parameters. For problems with large numbers of spectral index regions, a method such as Hamiltonian Monte Carlo \citep{DUANE1987216, hmcneal} could be used. By keeping to a relatively low number of regions, we are able to use the affine-invariant MCMC ensemble sampler {\tt emcee} package \citep{emcee}. The log-likelihood is given by
\begin{equation}
    \textrm{ln} \mathcal{L} = - \frac{1}{2} \sum_j \left(\dvec_j - \mvec_j \right)^{\rm T} \Nmtrx_j^{-1} \left(\dvec_j - \mvec_j\right),
\end{equation}
where the data model for a frequency channel labelled by $j$ is now given by
\begin{equation}
\mvec_j = \sum_k \Mmtrx_j \, \Theta_k \svec \left({\nu_j}/{\nu_0}\right)^{\beta_k}.
\end{equation}
Here, each spectral index region is labelled by $k$, and $\Theta_k$ is a pixel mask that is 1 for pixels within the region, and zero otherwise. Recall that the $\Mmtrx$ operator can include a beam convolution term; this will operate on the masked map, resulting in non-zero model values in some pixels outside region $k$ (particularly those close to the boundary of the region).

 \begin{figure*}
    \centering
    \includegraphics[width=0.49\linewidth]{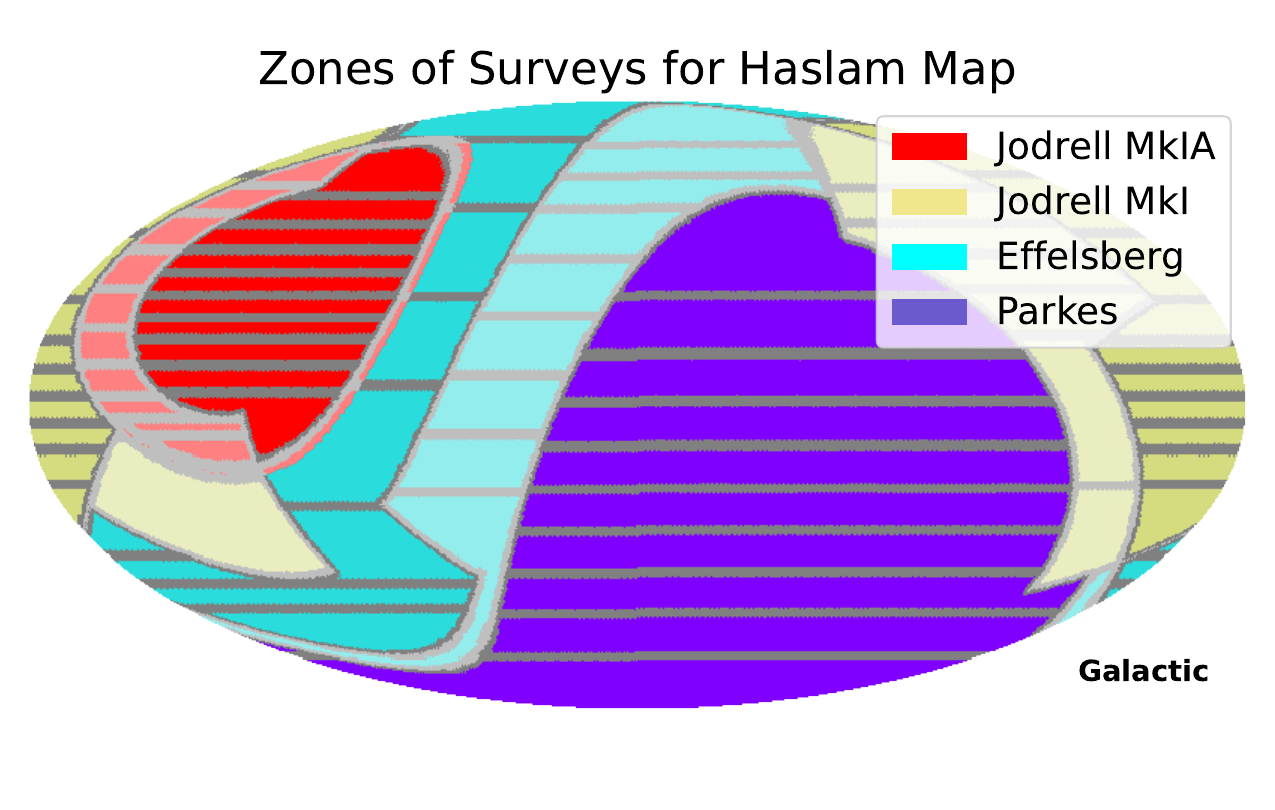}
    \includegraphics[width=0.49\linewidth]{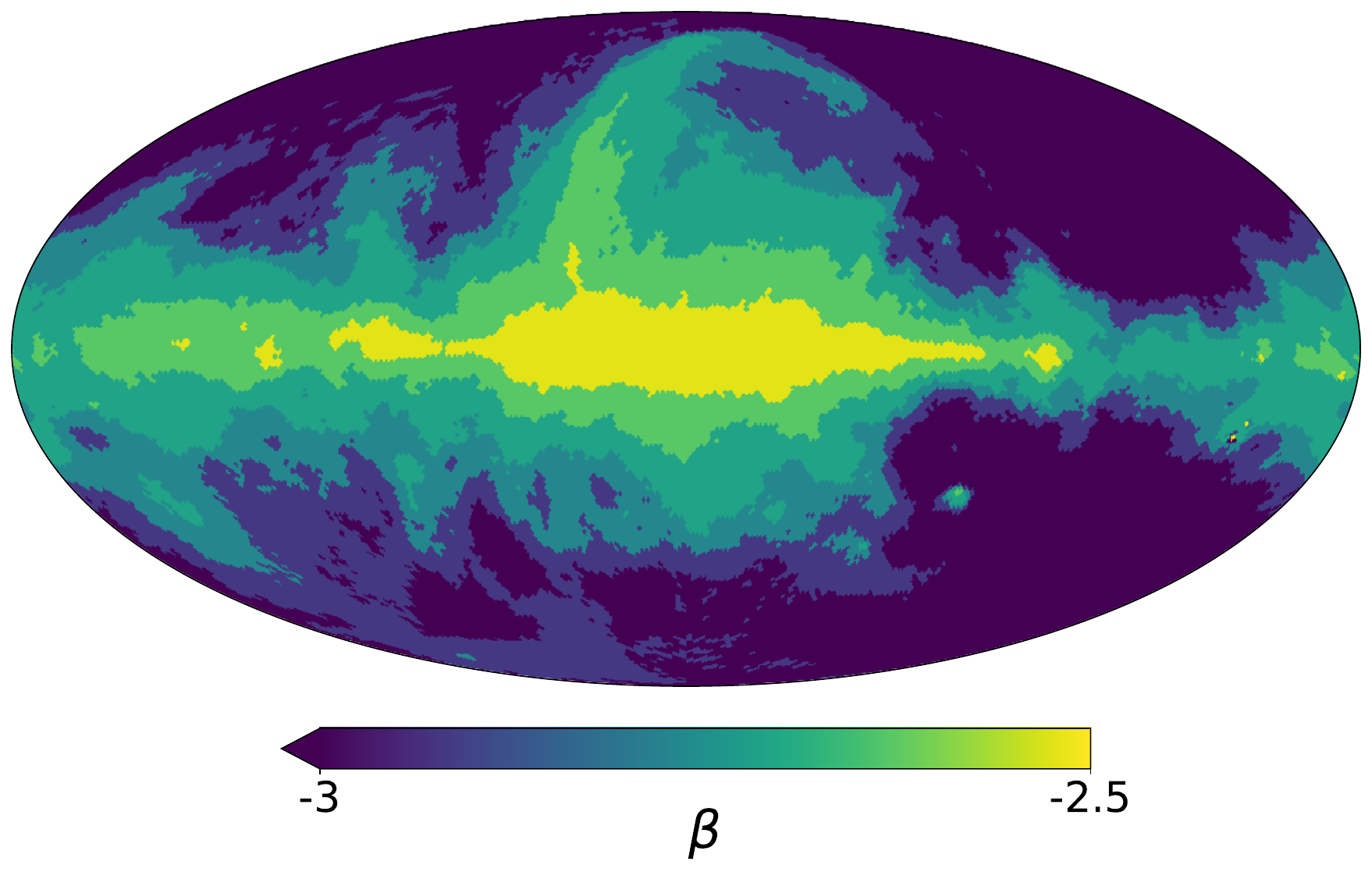}
    \caption{{\it(Left):} The different zones (in Galactic coordinates), corresponding to different surveys used to construct the Haslam map, taken from  \citet{haslam1982A&AS...47....1H}. We have combined the thin strip of overlap between the Jodrell Mk-IA and Effelsburg instruments, and further subdivided each zone into 10 smaller subzones represented by the grey outline. 
    {\it (Right):} The spectral index map used in the second part of the paper.  The `true' value of $\beta$ for each region is randomly drawn from a normal distribution i.e. $\beta \sim \mathcal{N}(\mu_\beta, 0.05)$, $\mu_\beta = [-2.5, -2.6, -2.7, -2.8, -2.9, -3]$ for each of the six zones.}
    \label{fig:haslam_zone}
    \label{fig:spectral_varying}
    
\end{figure*}

Different log-priors can be added to the likelihood expression to define the log-posterior function that is sampled from during the third step of each Gibbs iteration (see Eq.~\ref{eq:gibbsiter}). For non-linear parameters such as spectral indices, uniform priors are generally informative, and may cause the inferred spectral index parameter to be biased away from around its true value. This matter was discussed in detail by \cite{Eriksen_2008}, with the solution being to impose an `uninformative' Jeffreys prior to resolve the bias. We do not investigate the impact of non-uniform priors in this paper, but note this as an important consideration.

When sampling of $\boldsymbol{\beta}$ is enabled, this becomes the slowest part of each Gibbs iteration and significantly increases the run-time of the analysis. To speed it up, we avoid the need for expensive `burn-in' exploration of the parameter space by first performing a least-squares optimisation to find the maximum a posteriori solution for the conditional distribution (i.e. the third line of Eq.~\ref{eq:gibbsiter}). We then start the MCMC sampler from this point. After inspecting trace plots of the $\boldsymbol{\beta}$ parameters for a few longer MCMC chains, we found that the MCMC walkers appear to be reasonably well converged after performing this procedure.

To ensure an independent sample of $\boldsymbol{\beta}$ is drawn at each Gibbs iteration (i.e. not just the MAP solution), we then run the MCMC for 5 samples, and select the last one to adopt as the new draw of $\boldsymbol{\beta}$. More sophisticated monitoring of convergence would be possible here, but we retain this simple approach of using a fixed-length chain at each Gibbs step for the time being.

Moreover, to ensure that the MCMC of $\boldsymbol{\beta}$ is robust, we can adopt a more stringent prior on $\boldsymbol{s}$ by including the prior mean, $\boldsymbol{s}_0$, in the second Gibbs sampling step, which is now given by
\begin{equation}
   \left (\Xmtrx^{\rm T} \Nmtrx^{-1}\Xmtrx + \Smtrx^{-1}\right ) {\svec} = \Xmtrx^{\rm T}\Nmtrx^{-1}\dvec + \Xmtrx^{\rm T} \Nmtrx^{-1/2}\omega_n + \Smtrx^{-1/2}\omega_{s} + \Smtrx^{-1} \svec_0. \label{eq:gcr_s_prior}
\end{equation}
For clarity, the additional term $\Smtrx^{-1} \svec_0$ arises from the Gaussian prior likelihood being $\propto \exp \left (-\frac{1}{2} (\svec - \svec_0)^{\rm T} \Smtrx^{-1}(\svec - \svec_0) \right )$.

\section{Simulations of the observed sky}
\label{sec:simulation}
In this section, we describe how we simulate the data described in the previous section, and define the priors used in the Gibbs sampling and MCMC process.

\subsection{Flux scale factor model}

We divide the sky into four zones according to the surveys used to produce the Haslam map, shown in the left panel of Figure~\ref{fig:haslam_zone}. In the simplest case, assuming constant flux scale factors per survey, there should only be four unique values in $\gvec$ in total, where in the overlapping areas represented by the different colour shades, the effective $\gvec$ should be based on a weighted average of the original four values. It is unclear how this dependency should be modelled however, as an unspecified tapering of the edges of regions was performed in the original \citet{haslam1981A&A...100..209H} paper. To guard against making incorrect assumptions, we promote the overlapping areas to regions with separate flux scale factors in their own right. In the absence of more specific knowledge about the actual Haslam flux scale factors, we generate a notional set of $\gvec$ values for each zone by randomly drawing $\delta \gvec \sim \mathcal{N}(0, 0.05)$, where we recall that $\mathbfit{1}$ is added to the pixel-space representation of the flux scale factor map $\gvec$ so that the `default' scale factor would be unity (corresponding to perfect calibration). This is a conservative choice for the flux scale factor values, to match the value assumed by \citet{remazeilles_10.1093/mnras/stv1274}; as discussed above, significantly larger values may in fact apply.

To avoid complications, we have also assumed that the thin strip of overlap between the Jodrell Mk-IA and Effelsburg telescopes has the same flux scale factor as the unique Jodrell Mk-IA zone. Next, we divide each zone into 10 smaller subzones, so that the effective number of flux scale factor parameters is 70. 10 subzones is just an arbitrary number that we have decided to ensure that the flux scale factors are not overfitted. The pixels in the subzones are grouped together based solely by their index, such that each subzone would have $N_{\rm pix, zone}$/10 consecutive pixels with $N_{\rm pix, zone}$ being the total number of pixels in the entire zone. With this method of modelling the flux scale factors, the projection matrix $\Pmtrx$ is constructed to map each $\gvec$ value to the relevant pixels on the sky in \textsc{HEALPix}'s \textsc{Ring}-ordering. Note that we have neglected zero-point offsets in this treatment, but these can be included directly in the GCR equation if desired, since they are another linear term in the data model. We choose to set the inverse prior covariance $\Gmtrx^{-1} \to 0$, which amounts to a uniform prior with infinite support. Different scenarios for the assumed flux scale factors could \citep[e.g. as in][]{wilensky2024bayesian} could be encoded in the analysis through an appropriate choice of $\Gmtrx$, but we do not pursue this further here. 

\subsection{Diffuse emission sky  model}
\label{sec:diffusemodel}

We take our version of the all-sky map at 408~MHz from \textsc{pyGDSM} \citep{pygdsm_2016ascl.soft03013P}, which uses a principal component analysis algorithm to find the best fit components and spectra across 29 sky maps between 10 MHz to 5 THz \citep{gsmimproved10.1093/mnras/stw2525}. It then outputs \textsc{HEALPix} maps at the user's chosen frequencies. Note that although \textsc{pyGDSM} has the ability to generate random spectral indices $\beta$ following their best-fit model, we do not make use of this feature. Instead, we generate the high-resolution sky map at 408~MHz and subtract a CMB monopole of 2.725 K from it, before using \textsc{HEALpy}\footnote{\url{http://healpix.sourceforge.net}} \citep{2005ApJ...622..759G, Zonca2019} to degrade the map to $\textsc{nside}=64$, corresponding to an angular pixel area of 0.84 square degrees. Recall that no beam convolution is applied to this map in our model.

Next,  we generate sky maps between 50 and 150~MHz with $\Delta \nu=5$ MHz using Eq.~\ref{eqn:spectra}, with two sets of $\beta$ values: (a) $\beta = -2.52$ for all pixels, corresponding to the best-fit mean reported by \cite{gsmimproved10.1093/mnras/stw2525}; and (b) six zones of $\beta$, similar to \cite{anstey10.1093/mnras/stab1765}, in which the value of $\beta$ for each region is randomly drawn from a normal distribution $\beta \sim \mathcal{N}(\mu_\beta, 0.05)$ where $\mu_\beta = [-2.5, -2.6, -2.7, -2.8, -2.9, -3]$ respectively. A map of the values of $\beta$ is shown in the right panel of Figure \ref{fig:haslam_zone}.

After generating the temperature maps at lower frequencies, instead of following the order outlined in Equation \ref{eqn:data}, we switch the order of the smoothing and downgrading process. Specifically, we degrade them down to a pixel grid of $\textsc{nside}=16$ or $\textsc{nside}=8$ before convolving with a beam and adding noise, both of which are outlined in the next section. Note that we degrade before convolving for computational performance reasons; the map resolutions are still sufficient to permit accurate beam convolutions given the large beam width. For more complex beams with finer angular structure, the order of operations would need to be reversed, with a corresponding reduction in performance.

We have also considered both all-sky and incomplete sky coverage. For the incomplete sky coverage, we have made the assumption that the unobserved sky region in the lower frequency maps is at declinations $\leq - 60^\circ$.
%completely coincides with the Jodrell Mk-IA region with declination $\geq 45^\circ$. This is to avoid complications that could arise from the mis-match of pixels included in each flux scale factor zone, which could be achieved with real data through appropriate flagging.
In addition, we have set the prior variance of the true signal to be $10 \%$ of the signal values i.e. $\Smtrx= (0.1 \times \svec )^2$ in each pixel, with no assumed correlations between pixels. In the varying spectral indices case, we have also included a prior mean $\svec_0$, with values drawn from a Gaussian distribution centred on the true flux values $\svec_{\rm true}$, with a covariance of $\Smtrx$ i.e. $\svec_0 \sim \mathcal{N} (\svec_{\rm true}, \Smtrx)$.

\subsection{Beam and noise model}

The sensitivity of a radio telescope can be calculated from the radiometer equation, which (in temperature units) is given by
\begin{equation}
    \sigma_{\rm noise} = \frac{ T_{\rm sys}}{ \sqrt{\Delta\nu \Delta t_{\rm p}}},
\end{equation}
where $T_{\rm sys}$ is the system temperature usually given by the sum of the sky and instrument temperatures, i.e. $T_{\rm sys}= T_{\rm sky} + T_{\rm inst}$ , $\Delta\nu$ is the frequency channel width, and $ \Delta t_{\rm p}$ is the integration time per pointing or volume element of the data. At low frequencies $\nu \lesssim 300$ MHz, $T_{\rm sys}$ is generally dominated by $T_{\rm sky}$.

Making use of this fact, we construct a simple model for the data provided by a 21cm global signal experiment operating at low frequency by rescaling a `representative' noise level based on the EDGES experiment of around 25~mK in 390.6~kHz frequency channels \citep{edges2018Natur.555...67B}. For our tests, we assume that the data have been binned into coarser $\Delta\nu=5$~MHz frequency channels between 50 -- 150~MHz, resulting in 20 frequency channels. Assuming a 30$^\circ$ beam FWHM, a pixel grid of \textsc{nside}=8 (pixels with approximately $7.3^\circ$ sides) is sufficient to make well-sampled maps. The corresponding noise rms per pixel is then $\approx 50$~mK, assuming that the data were taken through drift scan observations within a constant-declination stripe on the sky with height given by the beam FWHM, and width corresponding to about 6 hours of local sidereal time.

We further assume that a number of such stripes have been observed uniformly by similar telescopes around the world to form full-sky maps with homogeneous noise properties \citep[c.f.][]{2025arXiv250621258I}. The noise rms is assumed to be independent of frequency and location. The beam is assumed to be a simple Gaussian, and is also independent of frequency, as is sometimes achieved with real data by deconvolving a beam estimate for each dataset before reconvolving with a broader, simpler Gaussian beam. Importantly, we also assume the flux scale of these observations to be perfectly calibrated, so that they can be used as a reliable reference.

\begin{figure*}
    \centering
    \includegraphics[width=0.9\linewidth]{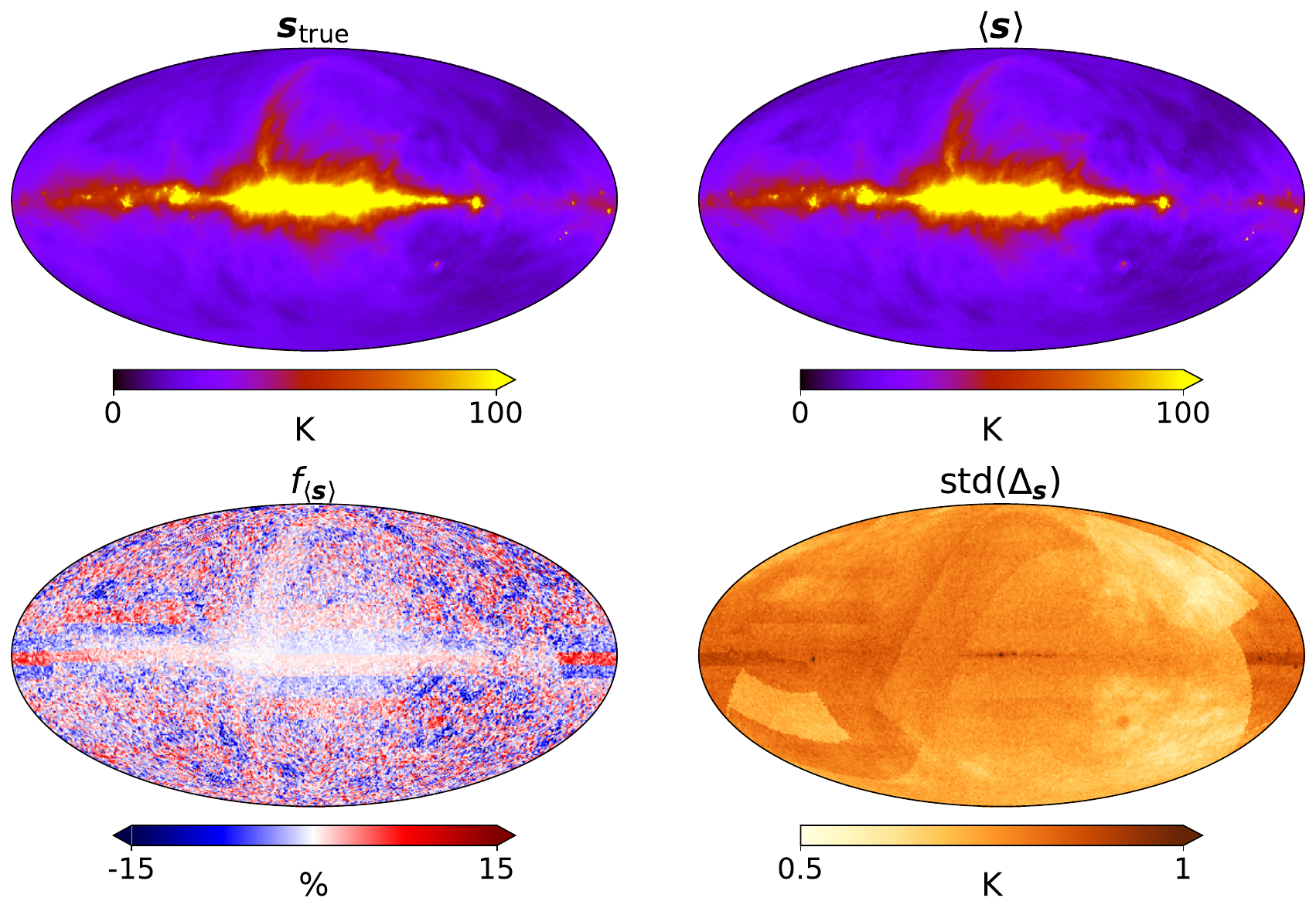}
    \caption{Clockwise from top left: the true brightness temperature map $\boldsymbol{s}$, the posterior mean of the sky temperature $\langle \boldsymbol{s}\rangle$, the standard deviation of the difference between the sample and true map, std($\Delta_{\boldsymbol{s}}$), and the fractional difference between $\langle \boldsymbol{s}\rangle$ and $\boldsymbol{s}$, $f_{\langle \boldsymbol{s}\rangle}$ for the {\sc STD} case.}
    \label{fig:s_results}

    \centering
    \includegraphics[width=0.9\linewidth]{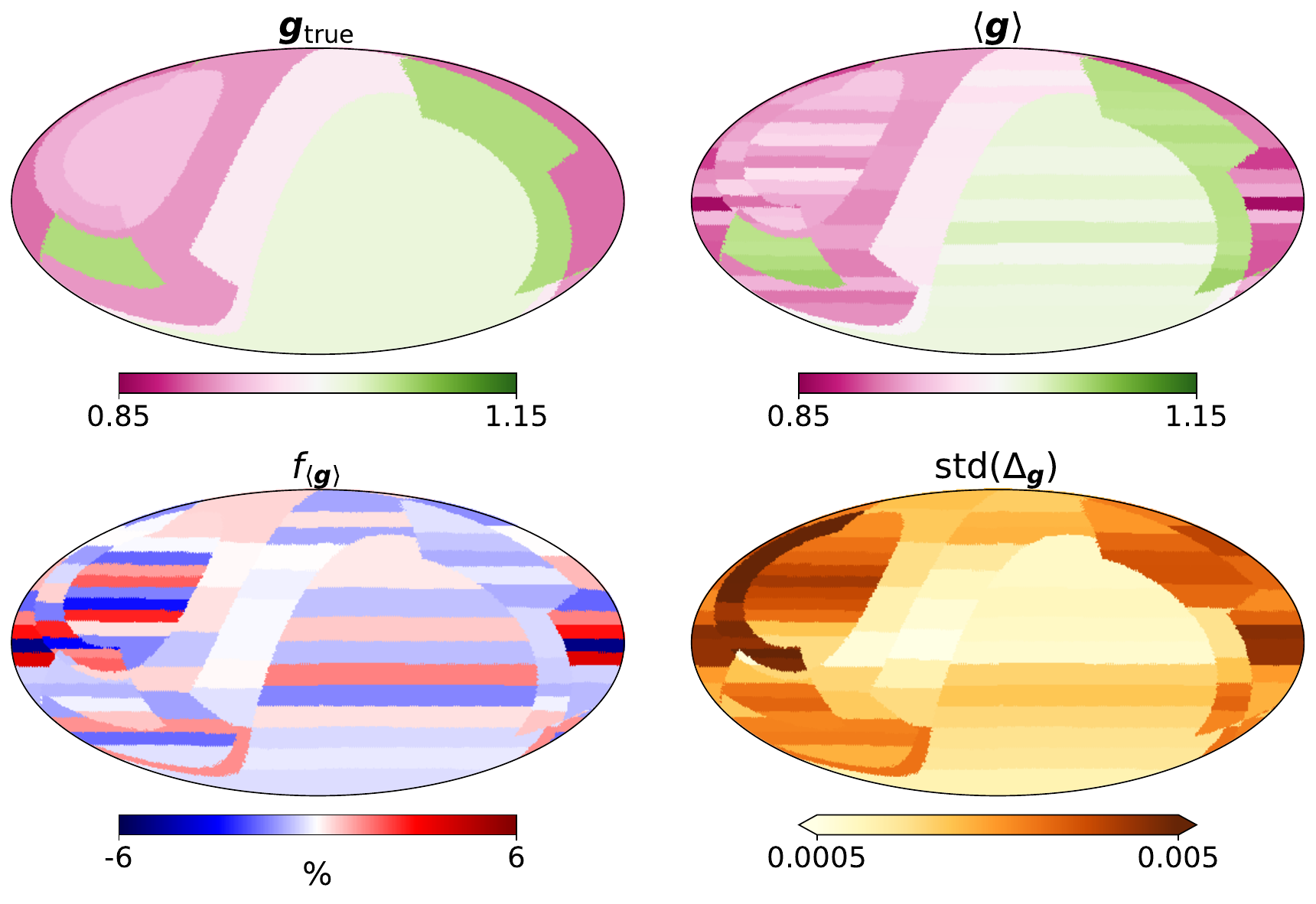}
    \caption{Clockwise from top left: the true flux scale factor $\boldsymbol{g}$, the estimated flux scale factor $\langle \boldsymbol{g}\rangle$, the standard deviation of their difference std($\Delta_{\boldsymbol{g}}$), and their fractional difference $f_{\langle \boldsymbol{g}\rangle}$ for the {\sc STD} case.}
    \label{fig:g_results}
    
\end{figure*}

These choices have been made for the sake of simplicity, and there is no fundamental limitation of our method with respect to any of them. More sophisticated beam models can readily be included, although beam convolutions with beams that are non-axisymmetric can be considerably slower. Inhomogeneous and frequency-dependent noise variance can be included directly, although non-white noise (i.e. off-diagonal entries in the noise covariance matrix) make the inverse covariance more numerically demanding to calculate. Similarly, increasing the spectral and spatial resolution of the data is possible, but will result in slower run-times unless suitable optimisations or parallel processing are implemented. Implementing flux scale uncertainties for the global signal experiment is also possible, following the same maths as for the $\gvec$ parameters applied to the Haslam map (see Sect.~\ref{sec:gcr}), although well-specified priors would be required to prevent substantial degeneracies from arising.

The `ground truth' model for the Haslam map was already described in Sect.~\ref{sec:diffusemodel}. We do not include an explicit beam model in this, as per Eq.~\ref{eq:dH}, although one could be included straightforwardly; instead, we assume that the Haslam data are at their `natural' resolution, and beam effects for this dataset can be ignored. It is also unclear what the actual noise rms of the Haslam map is, and so we have considered two values that have been used in the literature \citep[e.g.][]{remazeilles_10.1093/mnras/stv1274}: 800 and 1300~mK per pixel. For the latter, we investigate the effects of having an incorrect assumption of the noise level on the parameter constraints in Sect.~\ref{sec:fixed_beta}. We do not consider spatially-varying noise or noise correlations due to (e.g.) residual $1/f$ noise. However for future work, it would be interesting to include noise parameters in the inference as well.

\begin{figure}
    \includegraphics[width=0.98\linewidth]{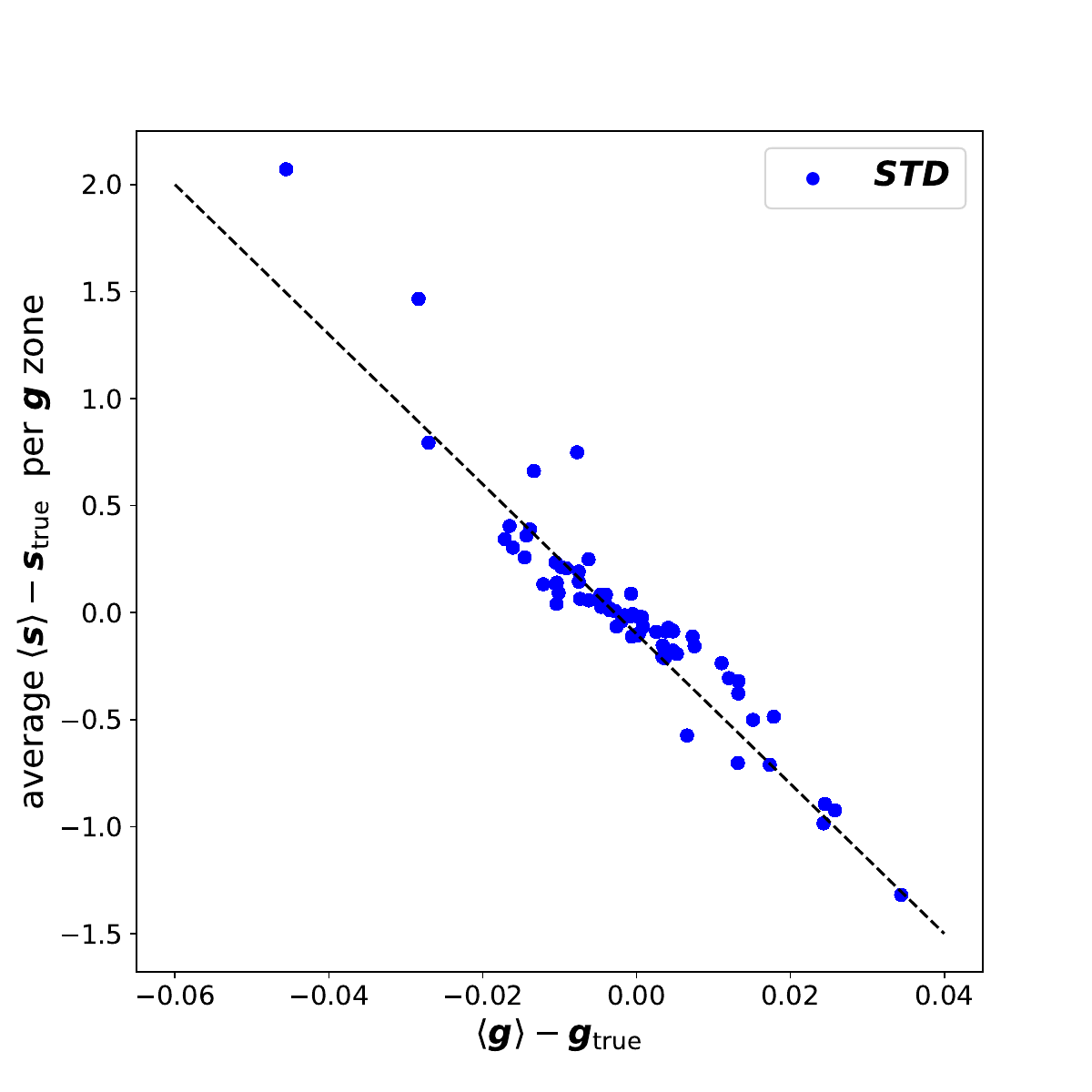}
    \caption{The average difference between $\langle \boldsymbol{s} \rangle$ and $\boldsymbol{s}_{\rm true}$ in each $\boldsymbol{g}$ zone with respect to the difference between $\langle \boldsymbol{g} \rangle$ and $\boldsymbol{g}_{\rm true}$ for the {\sc STD} case. The values of $\svec$ are in Kelvin. A total of 70 zones are plotted. The dashed line shows the line of best-fit, constrained to pass through the origin.}
    \label{fig:avgs_vs_g}
\end{figure}

% \begin{figure*}
%     \centering
%     \includegraphics[width=0.95\linewidth]{figures/avg_s_vs_g.pdf}
%     \caption{The average difference between $\langle\boldsymbol{s} \rangle$ and $\boldsymbol{s}_{\rm{true}}$ in each $\boldsymbol{g}$ zone (left), and the difference between $\langle\boldsymbol{g} \rangle$ and $\boldsymbol{g}_{\rm{true}}$ (right) for the {\sc STD} case with fixed $\beta$. There is a clear anti-correlation between the two parameters.}
%     \label{fig:avgs_vs_g}
% \end{figure*}

\section{Results}
\label{sec:results}
In this section, we present our findings on the constraints with the different spectral models, beam sizes, and noise levels. We start by considering a fixed spectral index and full sky coverage in Section \ref{sec:fixed_beta}, before extending our analysis to limited sky coverage and regions of varying spectral indices in Sections \ref{sec:limited_coverage} and \ref{sec:varying_beta} respectively. For clarity, we obtain the posterior mean of the sky temperature, flux scale factor, and spectral indices in each pixel/zone by averaging over the 1000 samples returned for each case by the Gibbs sampler, and hereafter, we refer to them as $\langle \boldsymbol{s}\rangle$, $\langle\boldsymbol{g}\rangle$, and $\langle\boldsymbol{\beta}\rangle$ respectively. 

For each case, we calculate the integrated auto-correlation time, $\tau = \sum_t C(t)/C(0) $, where $C(t)$ is the auto-covariance function at lag $t$ \citep{goodmanweare2010CAMCS...5...65G}. When $\beta$ is fixed, the mean $\tau$ across all pixels or zones for parameters $\boldsymbol{s}$ and $\boldsymbol{g}$ ranges between 3 -- 26 and 2 -- 11 respectively. The range of the mean effective sample size, which in our case is given by $1000/\tau$, is thus  38 -- 345 and 88 -- 468 for $\boldsymbol{s}$ and $\boldsymbol{g}$ respectively. When $\beta$ varies, however, the mean $\tau$ for $\boldsymbol{s}, \boldsymbol{g},$ and $\beta$ is around 112, 22, and 72 respectively, corresponding to an effective sample size of 9, 44, and 14, which is rather low. As a practical matter, it would be valuable to find alternative parametrisations or sampling methods to avoid the long chain correlation lengths caused by strong correlations between parameters when $\beta$ is allowed to vary, although we do not pursue this further in this work. As a further note, the chains for $\beta$ were manually inspected for convergence, in accordance with the description in Section \ref{sec:mcmc_beta}.
For all cases presented below, we have access to the full joint posterior distribution of almost 50,000 parameters: values of $\svec$ in 49,152 pixels, values of $\gvec$ in 70 regions, and when enabled, values of $\beta$ in six zones.

% \begin{figure*}
%     \centering
%     \includegraphics[width=\linewidth]{figures/estimates_nside_mask.pdf}
%     \caption{Maps of the fractional (left panels) and standard deviation of the difference (right panels) for the sky brightness temperature field (top row) and flux scale factors (bottom row) when sky coverage is incomplete at declination $\leq -60^\circ$.}
%     \label{fig:sg_map_mask}
% \end{figure*}

\subsection{Fixed spectral index}
\label{sec:fixed_beta}
For all cases considered in this section, we have set $\beta=-2.52$ for all pixels in the sky map.

\subsubsection{The standard case}
\label{sec:std_case}
We first look at the posterior mean of the sky signal and flux scale factors for our standard case ({\sc STD}), where the noise levels for the global experiment and Haslam map are assumed to be 50 and 800 mK per pixel respectively. We present a map of the posterior mean of the sky signal for case {\sc STD} in Figure  \ref{fig:s_results}. The panels show the true sky $\boldsymbol{s}_{\rm true}$ (upper left), the posterior mean of the sky temperature $\langle \boldsymbol{s}\rangle$ (upper right), the standard deviation of their difference std($\Delta_{\boldsymbol{s}}= \langle \boldsymbol{s}\rangle - \boldsymbol{s}_{\rm true}$) (lower right), and their fractional difference $f_{\langle \boldsymbol{s}\rangle} = \langle \boldsymbol{s}\rangle / \boldsymbol{s} - 1$ (lower left). We have limited the colour scale ranges of $\boldsymbol{s}$ and $\langle \boldsymbol{s}\rangle$ to 100~K, $f_{\langle \boldsymbol{s}\rangle}$ to $\pm 15 \%$, and std($\Delta_{\boldsymbol{s}}$) to be between 0.5 and 1~K.

The smallest std($\Delta_{\boldsymbol{s}}$) coincides with the coldest area in the sky, while the largest std($\Delta_{\boldsymbol{s}}$) is in Galactic plane and the North Polar Spur. For the $f_{\langle \boldsymbol{s}\rangle}$ however, the middle of the Galactic plane has the lowest values while the other regions have values that resemble white noise. The largest values are on the far edge of the Galactic plane, where imprints of the flux scale factor zones can be seen. Recall that a Gaussian prior on $\svec$ with a standard deviation of 10\% around the true values has been imposed.

We then plot the same quantities for the estimated flux scale factor in Figure \ref{fig:g_results}: the true flux scale factor $\boldsymbol{g}$ (upper left), the posterior mean of the flux scale factor $\langle \boldsymbol{g}\rangle$ (upper right), the standard deviation of their difference std($\Delta_{\boldsymbol{g}}$) (lower right), and their fractional difference $f_{\langle \boldsymbol{g}\rangle}$ (lower left). We have capped the plotted values of $\boldsymbol{g}$ and $\langle \boldsymbol{g}\rangle$ to (0.85, 1.15),  $f_{\langle \boldsymbol{g}\rangle}$ to $\pm 6 \%$, and std($\Delta_{\boldsymbol{g}}$) to $(5 \times 10^{-4}, 5 \times 10^{-3} )$. For both $f_{\langle \boldsymbol{g}\rangle}$ and std($\Delta_{\boldsymbol{g}}$), the values in all regions are equally low with $ f_{\langle \boldsymbol{g}\rangle} \sim 1-2 \%$ and std($\Delta_{\boldsymbol{g}}) \leq 1 \times 10^{-3}$, except for the far edge of the Galactic plane where $f_{\langle \boldsymbol{s}\rangle}$ values are large, as in the previous figure.

To properly establish the relationship between the two parameters, we calculate the average difference between $\langle\boldsymbol{s} \rangle$ and $\boldsymbol{s}_{\rm{true}}$ in each of the 70 flux scale factor zones, and plot them against the difference between $\langle\boldsymbol{g} \rangle$ and $\boldsymbol{g}_{\rm{true}}$ in Figure \ref{fig:avgs_vs_g}. This shows a clear anti-correlation between $\boldsymbol{g}$ and $\boldsymbol{s}$ that accounts for the slight biases in the marginal means of these parameters.

\begin{figure}
    \includegraphics[width=0.98\linewidth]{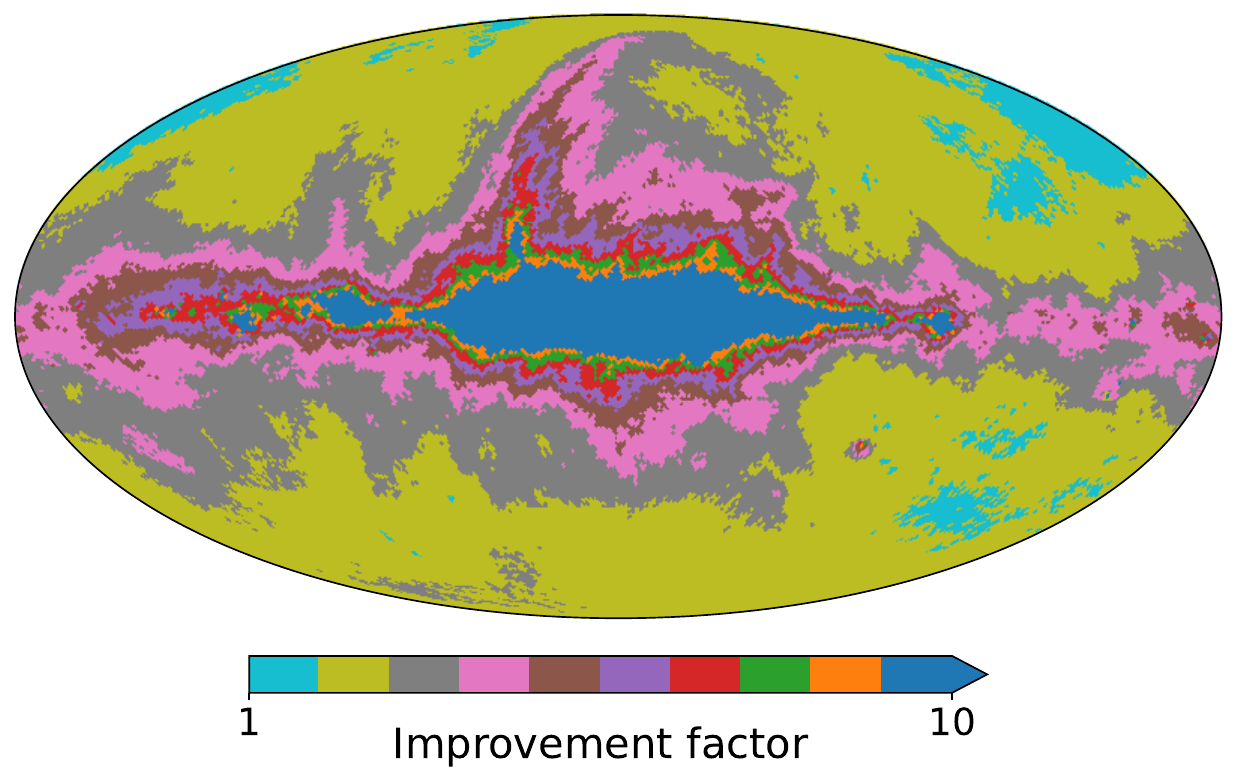}
    \caption{The `improvement factor' of the posterior mean of the temperature field, $\langle \svec \rangle$, compared with the prior on $\svec$, calculated as the ratio of the prior standard deviation (10$\%$ of $\svec_{\rm true}$) divided by the standard deviation of $\Delta_\svec$. A discrete colour map has been used to make it easier to identify the approximate improvement factor across the map, with cyan denoting the smallest improvement factor (with a value of 1 implying no improvement over the prior). The colour scale has been clipped at a value of 10, but values around 50 are achieved close to the Galactic centre.}
    \label{fig:improv_factor}
\end{figure}

Figure \ref{fig:improv_factor} shows the `improvement factor' on the recovered temperature field $\svec$ compared with the assumed prior. This is computed as the ratio of the prior standard deviation (set to 0.1 $\svec_{\rm true}$ in each pixel) to the standard deviation of $\Delta_\svec$ (the difference between the posterior mean and true values of $\svec$ in each pixel). This makes it clear that the recovered sky map is being improved (brought closer to the true value) everywhere by the addition of the low-frequency absolutely-calibrated data, despite its low resolution and also needing to marginalise over the flux scale factors. The improvement factor is greatest around the Galactic plane, with only a mild improvement at higher Galactic latitudes -- although this is a still a factor of a few everywhere but a small patch at quite high northern latitudes where the sky brightness is at its minimum.

\begin{figure*}
    \centering
    \includegraphics[width=0.9\linewidth]{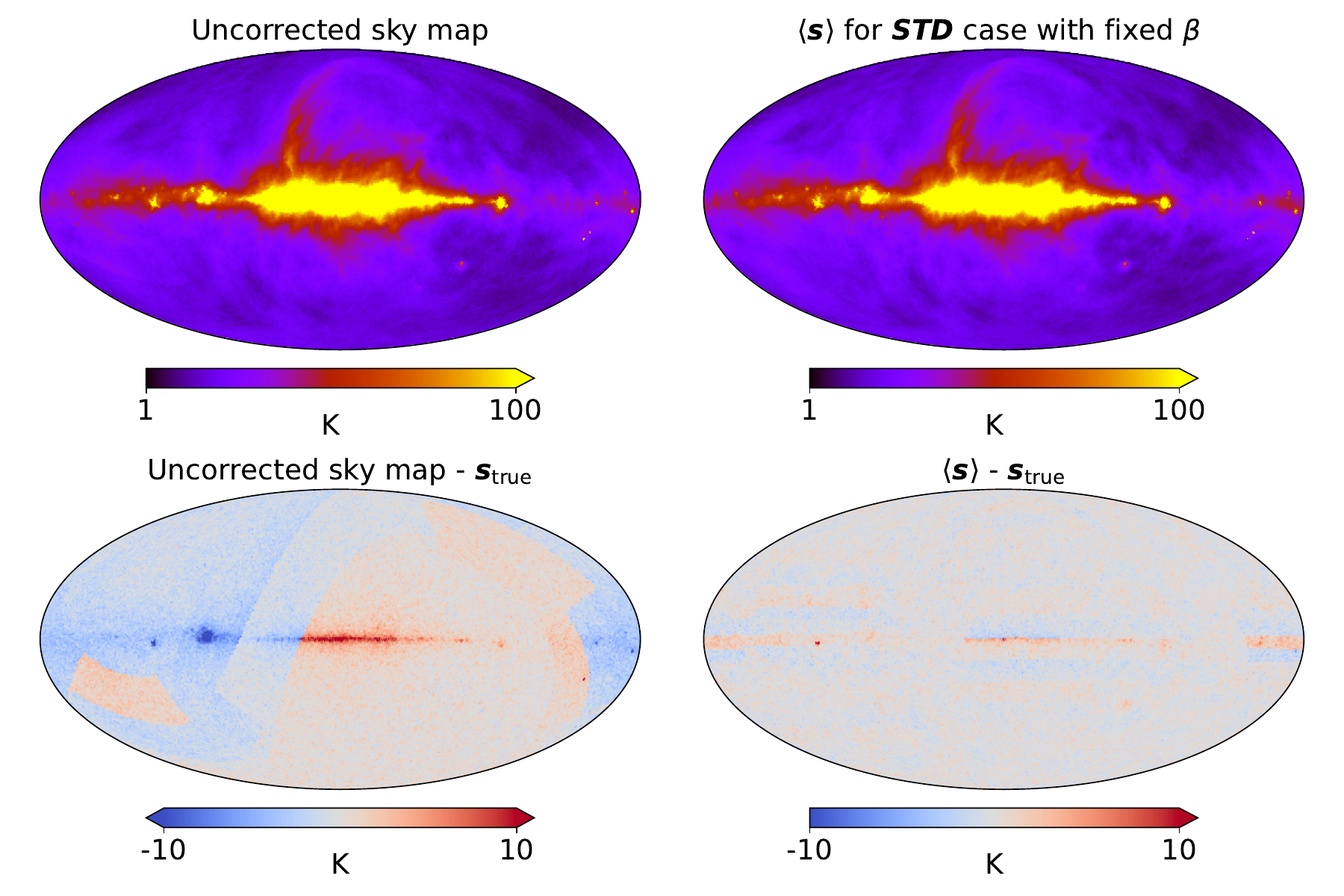}
    \caption{The sky map with uncorrected flux scale factors (top left) and the posterior mean (top right), along with their difference with the true sky brightness temperature field (bottom panels) for the {\sc STD} case with fixed $\beta$.}
    \label{fig:skymap_std}
\end{figure*}

Next, we investigate the effect of the bias on the recovered sky i.e the posterior mean of the sky signal $\langle \boldsymbol{s}\rangle$. We present the uncorrected sky map and the recovered corrected sky brightness temperature, along with their difference from the true brightness temperature field in Figure \ref{fig:skymap_std}. Without the corrected flux scale factors, the difference can be higher than 10~K in regions close to the Galactic plane. Despite the biases reported above, our framework is able to correct the sky to within 1~K in a majority of the sky pixels, while in the more biased zones, the pixel values are correct to within 5~K.

\begin{figure*}
    \centering
    \includegraphics[width=0.9\linewidth]{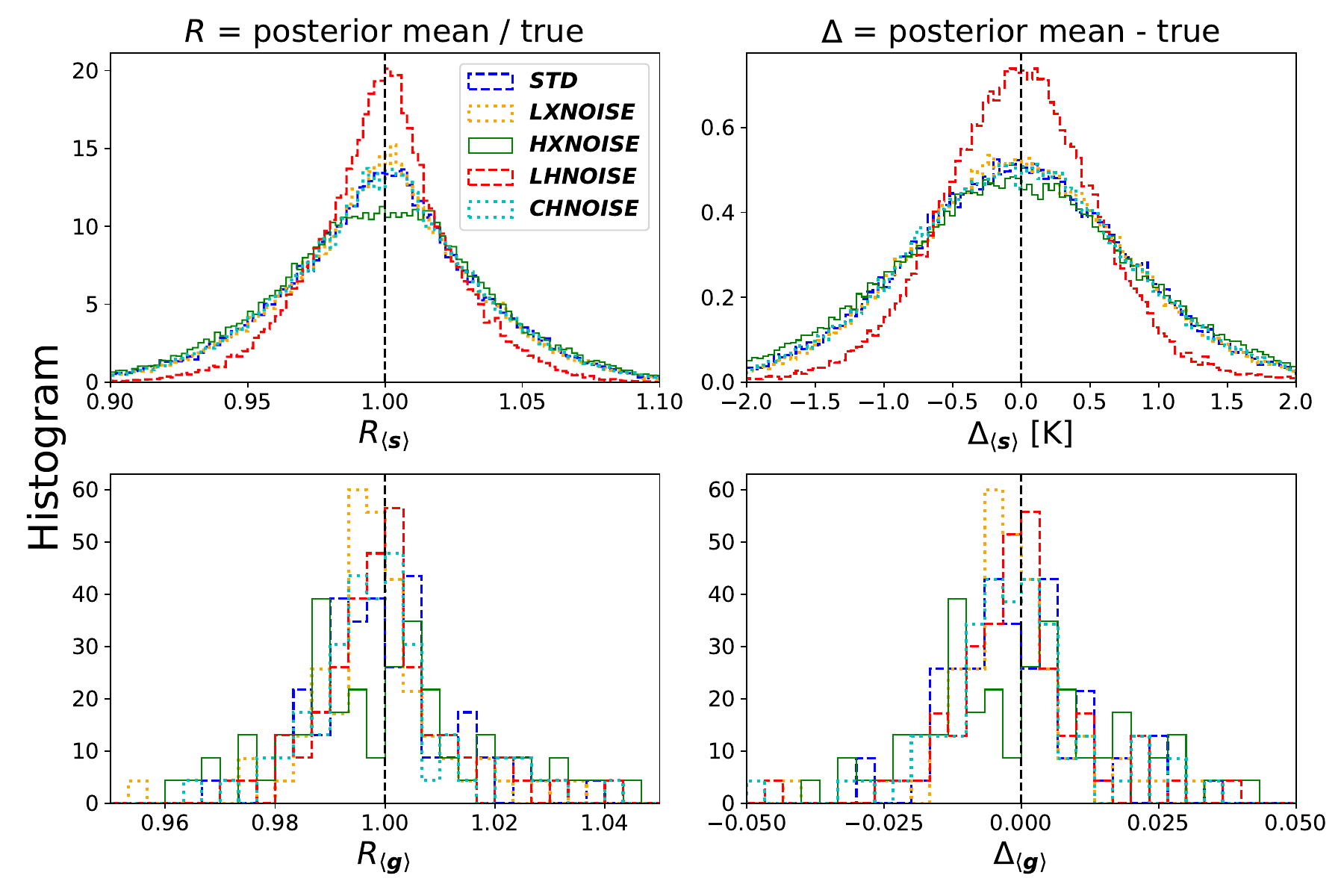}
    \caption{Histogram of the ratio (left panels) and difference (right panel) of the sampled sky temperature (top row) and flux scale factor (bottom row) to the true values for cases {\sc STD} (dashed blue), {\sc LXNOISE} (dotted orange), {\sc HXNOISE} (solid green), {\sc LHNOISE} (dashed red), and {\sc CHNOISE} (dotted cyan). The value of $\Delta_{\langle \boldsymbol{s} \rangle}$ is in Kelvin. The {\sc LHNOISE} case gives the tightest signal constraint.}
    \label{fig:noise_comparison}

\end{figure*}

\subsubsection{Different noise levels}

Next, we explore the effects of using different experimental assumptions on the constraints on the flux scale factor and true sky given a fixed spectral index $\beta$, following the Gibbs sampling steps in Section~\ref{sec:gibbs}. We present the parameters used for each case in Table \ref{tbl:param_values}. We have five cases in total: {\sc STD} is the standard set with the default parameters; {\sc LXNOISE} is performed with lower noise rms per pixel (25~mK) for the low-resolution experiment $\mathbfit{d}_x$; {\sc HXNOISE} is performed with higher noise rms per pixel (175~mK) for $\mathbfit{d}_x$; {\sc LHNOISE} is performed with lower noise (500 mK) for $\mathbfit{d}_{\rm H}$; and {\sc CHNOISE} is performed with `confused' noise for $\mathbfit{d}_{\rm H}$ where the true noise level is 1300 mK but we have ``mistakenly'' assumed it to be 800 mK.

We start by comparing the posterior means of the sampled sky temperature and flux scale factors, $ \langle \boldsymbol{s} \rangle$ and $\langle \boldsymbol{g} \rangle$, to the true values for the  {\sc STD} (dashed blue), {\sc LXNOISE} (dotted orange), {\sc HXNOISE} (solid green), {\sc LHNOISE} (dashed red), and {\sc CHNOISE} (dotted cyan) cases in Figure \ref{fig:noise_comparison}. We plot the ratio (left panels) and difference (right panel) of the posterior means of the sky temperature (upper row) and flux scale factor (lower row) to the true values as normalised histograms over pixels for $\boldsymbol{s}$ or zones for $\boldsymbol{g}$. In the top panels, we have capped the values of the ratio $R_{ \langle\boldsymbol{s}\rangle} = \langle\boldsymbol{s}\rangle/{\boldsymbol{s}_{\rm true}}$ and the difference $\Delta_{\langle\boldsymbol{s} \rangle} = \langle\boldsymbol{s}\rangle -\boldsymbol{s}_{\rm true}$ at (0.9, 1.1) and ($-$2.0, 2.0)~K respectively. In the bottom panels, we have capped the values of the ratio $R_{\langle \boldsymbol{g}\rangle} $ and difference $\Delta_{\langle\boldsymbol{g}\rangle}$ at (0.95, 1.05) and ($-$0.05, 0.05) respectively. 

As evident in the upper panels, the {\sc LHNOISE} case gives the tightest signal constraint with 1$\sigma$ values of $\pm 0.03$ and $\pm 0.76$~K for $R_{ \langle\boldsymbol{s}\rangle}$ and $ \Delta_{\langle\boldsymbol{s}\rangle}$ respectively. All other cases have similar constraints on $\langle \boldsymbol{s}\rangle$, with 1$\sigma$ values of $\pm 0.05$ and $\pm 1.1$~K for $R_{ \langle\boldsymbol{s}\rangle}$ and $ \Delta_{\langle\boldsymbol{s}\rangle}$ respectively.
For the flux scale factor $\gvec$ in the lower panels, all cases have similar distributions, with 1$\sigma$ values of $\approx \pm 0.01$ for both $R_{\langle\boldsymbol{g}\rangle}$ and $\Delta_{\langle \boldsymbol{g}\rangle}$. It is thus encouraging to see that even with the wrong assumption on the noise level of the original Haslam map, we are still able to constrain both the sky temperature field and flux scale factor.

\begin{figure}
    \centering
    \includegraphics[width=\linewidth]{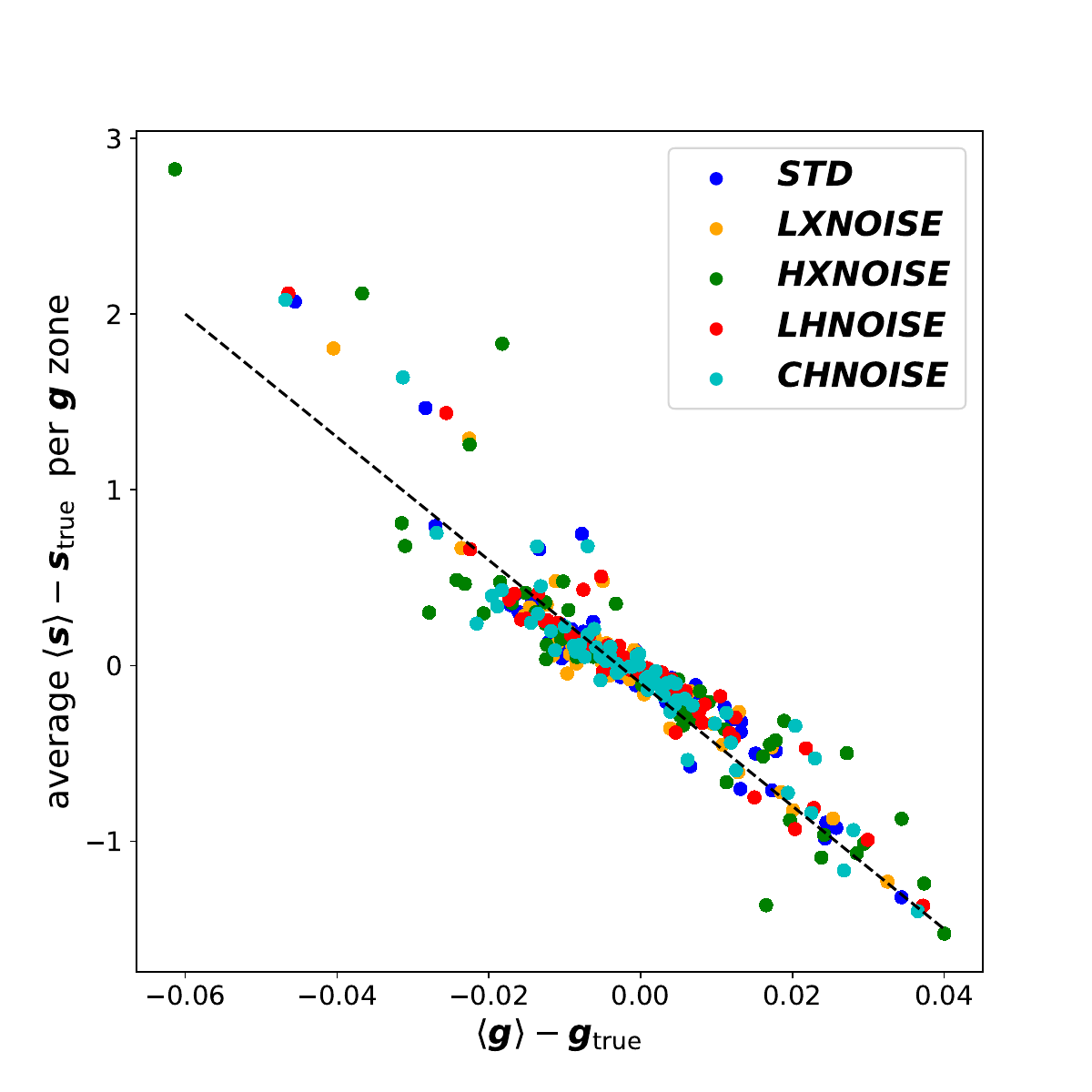}
    \caption{The average difference between $\langle \boldsymbol{s} \rangle$ and $\boldsymbol{s}_{\rm true}$ in each $\boldsymbol{g}$ zone with respect to the difference between $\langle \boldsymbol{g} \rangle$ and $\boldsymbol{g}_{\rm true}$ for the different cases.}
    \label{fig:scatter_s_vs_g}
\end{figure}

Finally, we plot the average difference between $\langle \boldsymbol{s} \rangle$ and $\boldsymbol{s}_{\rm true}$ in each $\boldsymbol{g}$ zone with respect to the difference between $\langle \boldsymbol{g} \rangle$ and $\boldsymbol{g}_{\rm true}$ for the different cases in Figure \ref{fig:scatter_s_vs_g}. As we have previously noted in Figure \ref{fig:avgs_vs_g}, the two parameters are anti-correlated. The {\sc HXNOISE} case gives the largest scatter compared to the other cases.

\begin{figure*}
    \centering
    \includegraphics[width=0.9\linewidth]{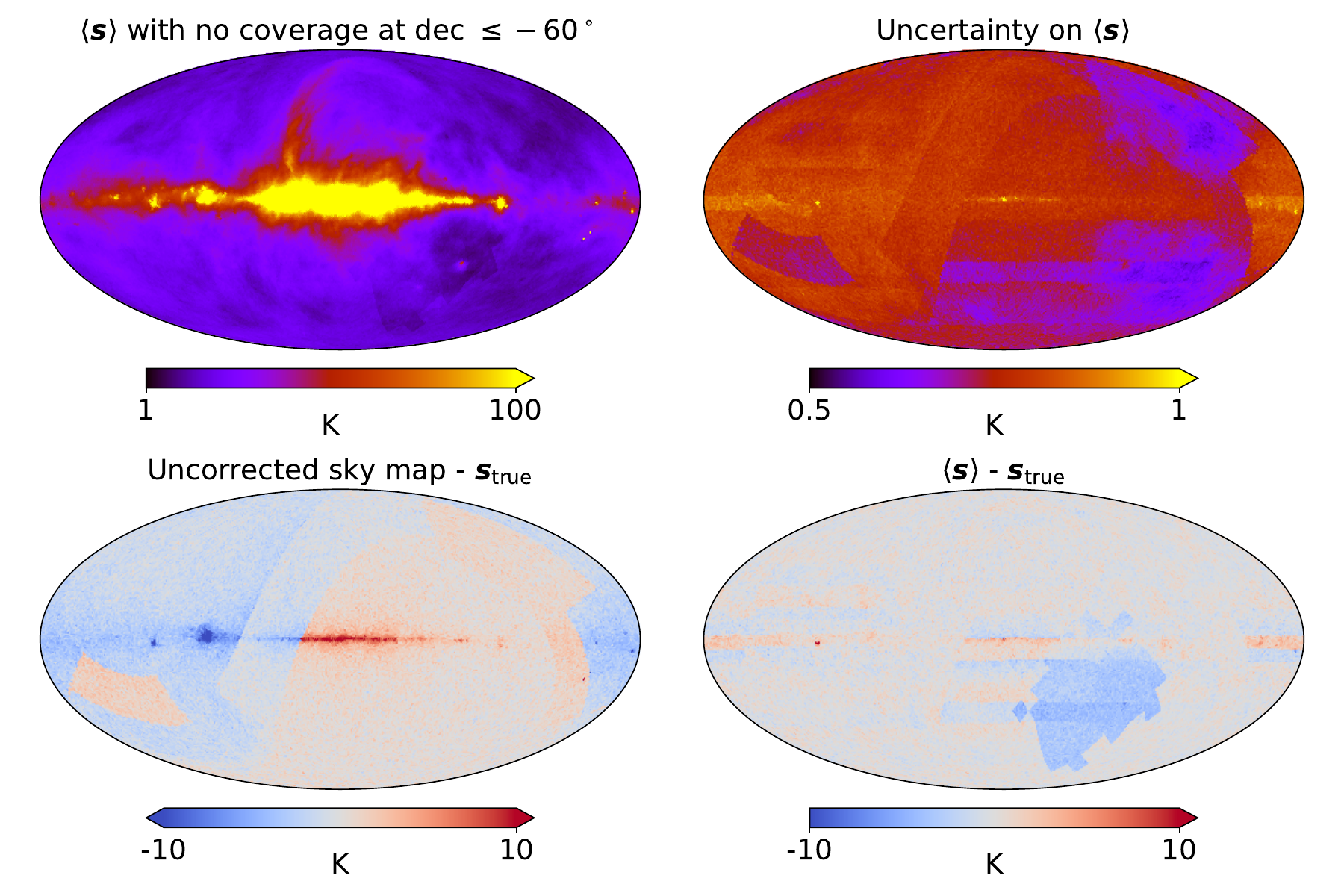}
    \caption{The posterior mean (top left) and the uncertainty (top right) of the sky brightness temperature for the {\sc STD} case with fixed $\beta$ and incomplete sky coverage at declination $\leq -60^\circ$, along with the difference between the uncorrected sky map (bottom left) and posterior mean (bottom right) with the true sky brightness temperature field.}
    \label{fig:skymap_mask}
\end{figure*}

\begin{figure*}
    \centering
    \includegraphics[width=0.95\linewidth]{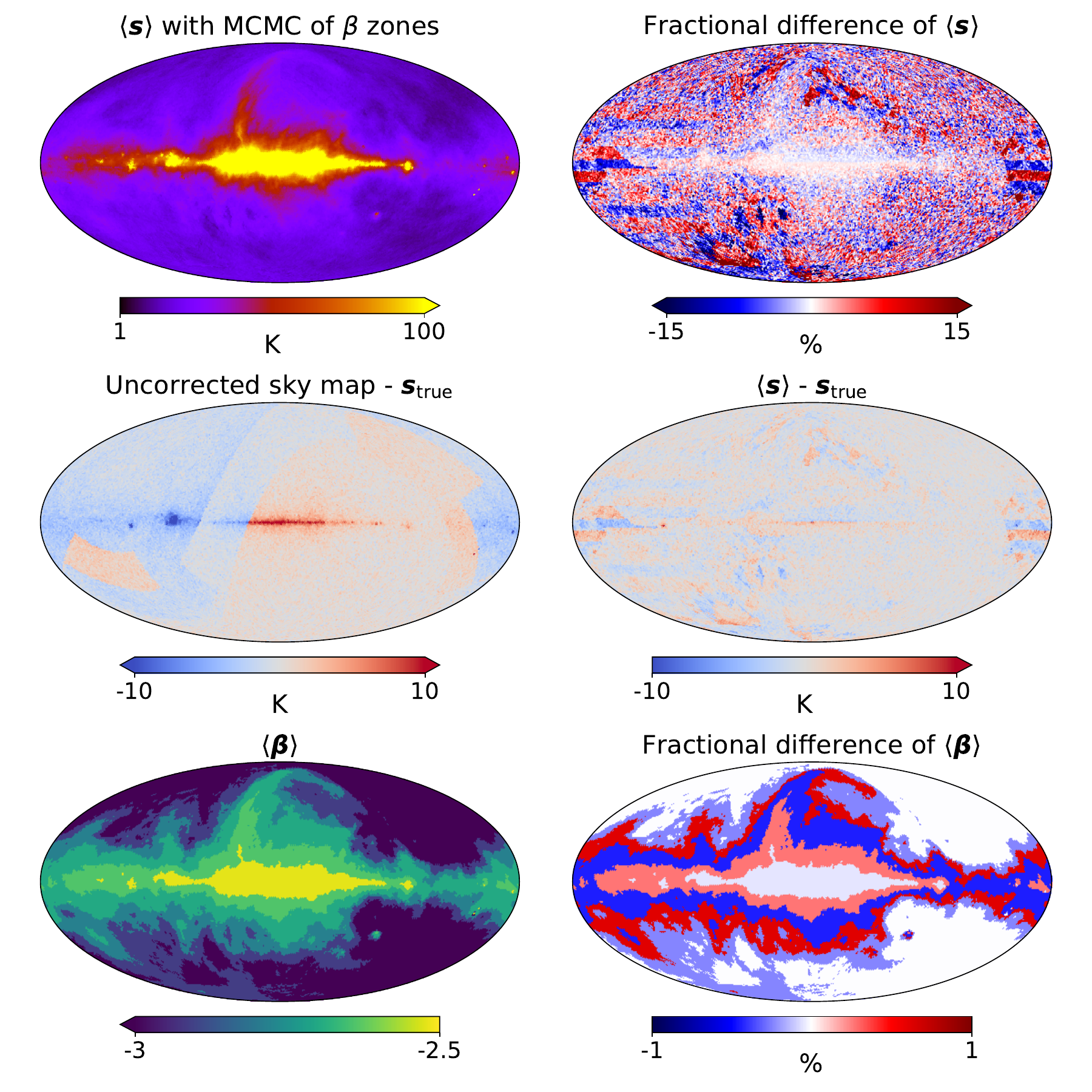}
    \caption{The posterior mean of the sky brightness temperature for the {\sc STD} case with MCMC of $\beta$ zones (top left) and its fractional difference with respect to the true map (top right), as well as the difference between the uncorrected sky map (middle left) and posterior mean (middle right) with the true sky brightness temperature field, and the posterior mean of the spectral index $\langle \boldsymbol{\beta}\rangle$ (bottom left) and its fractional difference with respect to the true values (bottom right).}
    \label{fig:skymap_spect}
\end{figure*}

% \begin{figure*}
%     \centering
%     \includegraphics[width=\linewidth]{figures/snr_sigma.pdf}
%     \caption{Maps of the SNR of the true sky, $\boldsymbol{s}/\sigma_{\rm noise}$, where we have assumed the noise level in the Haslam map to be 800 mK (left), and the ratio of the true sky to the standard deviation of the sampled sky, $\boldsymbol{s}/\sigma_{\langle \boldsymbol{s}\rangle}$ with varying $\beta$ and parameters from the {\sc STD} case (right).}
%     \label{fig:snr_sigma}
% \end{figure*}

\begin{figure*}
    \centering
    \includegraphics[width=0.49\linewidth]{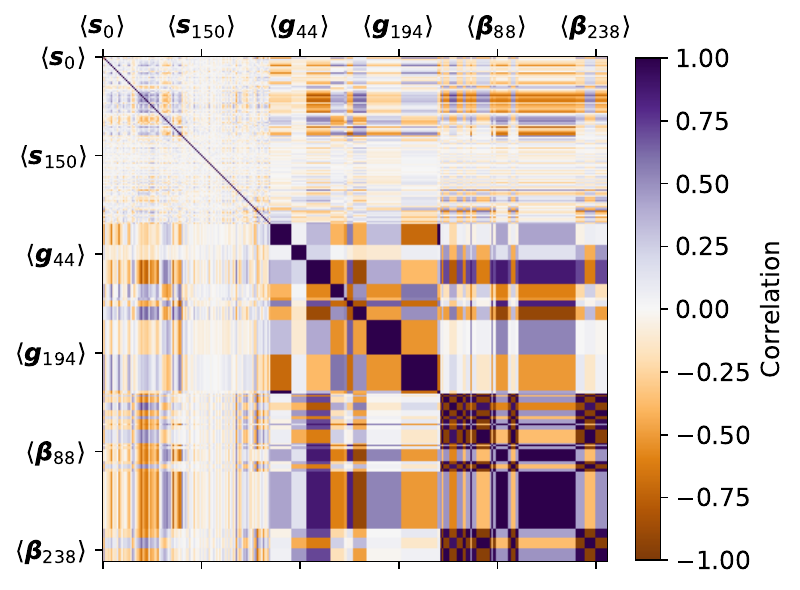}
    \includegraphics[width=0.49\linewidth]{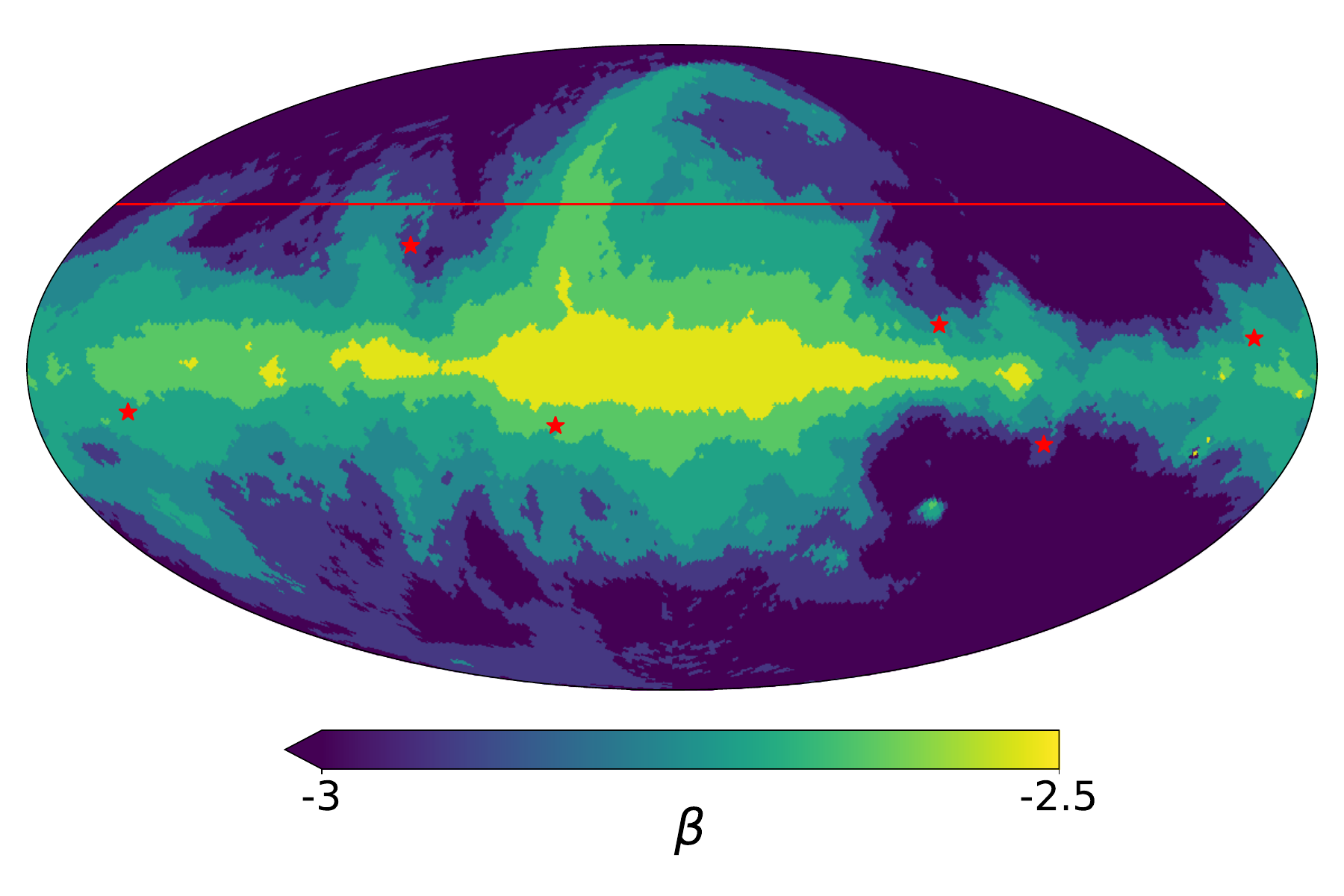}
    \caption{{\it(Left):} The correlation matrix of $\boldsymbol{s}, \boldsymbol{g}$, and $\boldsymbol{\beta}$ for a strip of 1024 consecutive pixels at constant Galactic latitude. This is computed from the covariance matrix, obtained by averaging over all 1000 posterior samples. {\it(Right):} Map showing the spectral zones, with the randomly-selected pixels used in Figure \ref{fig:sed} marked as red stars. The red line shows the strip of 1024 pixels used to evaluate the correlation matrix in the left panel.}
    \label{fig:}
    
    % The z-score, $z = (\boldsymbol{s} - \mu) / \sigma$ for the sky signal $\langle \boldsymbol{s}\rangle$.}
    \label{fig:corr_mtrx_zscore}
\end{figure*}

\begin{figure*}
    \centering
    \includegraphics[width=0.95\linewidth]{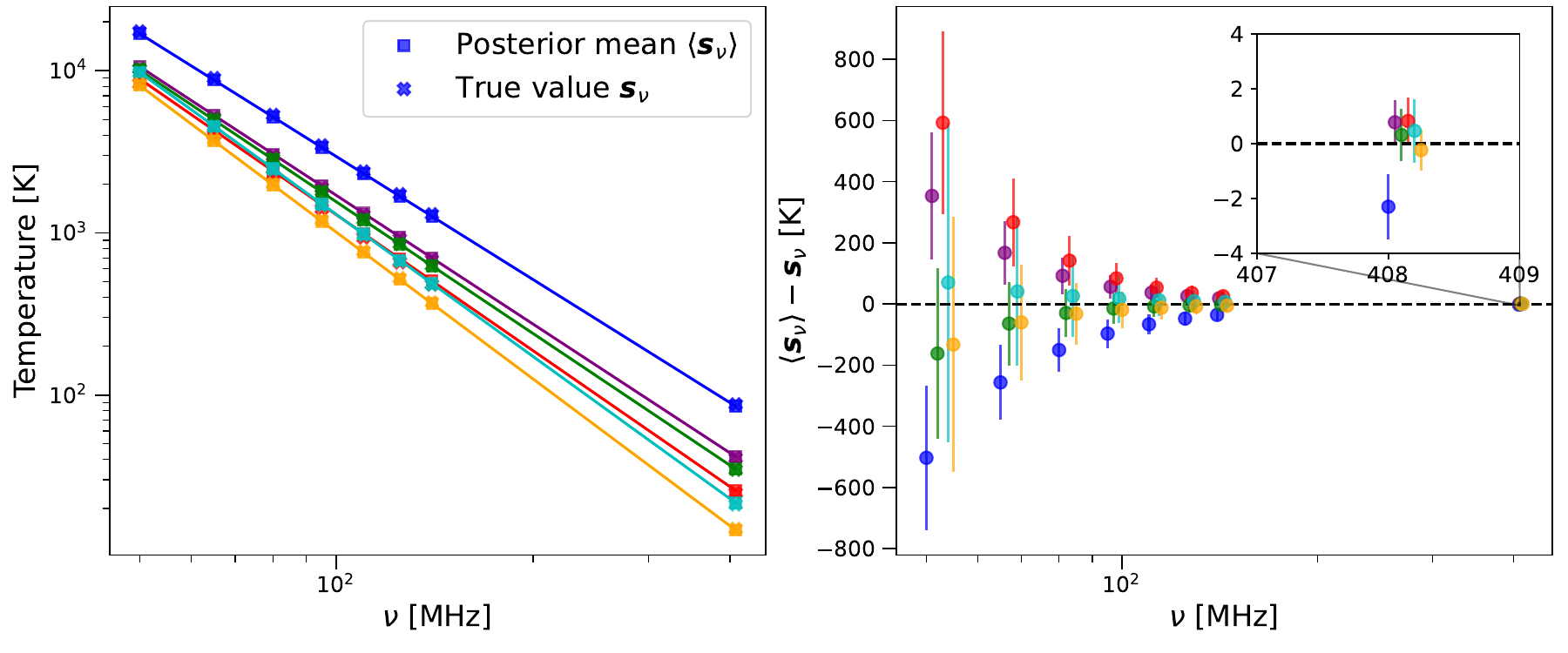}
    \caption{{\it(Left):} The mean of the posterior predictive distribution $\langle \boldsymbol{s}_\nu\rangle$ (square) and true $ \boldsymbol{s}_\nu$(cross) spectral energy distributions of six random pixels, one from each spectral zone, for six frequency channels. The chosen pixels are marked on the right-hand panel of Fig.~\ref{fig:corr_mtrx_zscore}. {\it(Right):} The mean and standard deviation of the difference (residual) between these quantities. The values on the x-axis have been slightly offset from the true values for clarity.}
    \label{fig:sed}
\end{figure*}

% \begin{figure}
%     \centering
%     \includegraphics[width=\linewidth]{figures/stars_lines.pdf}
%     \caption{Map showing the spectral zones, with the randomly-selected pixels used in Figure \ref{fig:sed} marked as red stars. The black dotted line shows the strip of 1024 pixels used to evaluate the correlation matrix in Figure~\ref{fig:corr_mtrx_zscore}.}
%     \label{fig:stars_line}
% \end{figure}

\subsection{Incomplete sky coverage}
\label{sec:limited_coverage}
Next, we investigate the effects of incomplete sky coverage on the constraints on the sky brightness temperature and flux scale factors. In Figure \ref{fig:skymap_mask}, we present maps of the posterior mean $\langle \boldsymbol{s}\rangle$ (top left) and the uncertainty (top right) of the sky brightness temperature for the {\sc STD} case with fixed $\beta$, now with incomplete sky coverage at declinations $\leq -60^\circ$ for the low-frequency experiment. The difference between the uncorrected sky map (bottom left) and posterior mean (bottom right) with the true sky brightness temperature field are also shown.

The pixel values of $\langle \boldsymbol{s}\rangle$ are mostly similar to the results of the {\sc STD} case presented in Section \ref{sec:std_case}, except in the unobserved region where they are underestimated, as is evident from the light blue region in the bottom left panel. With our framework, we are able to correct the sky to within 5~K difference even in the unobserved region. This suggests that information from surrounding regions is helping to constrain these pixels somewhat, albeit relatively ineffectively.

This is obviously dependent on the parametrisation; the long, thin flux scale factor zones in the Parkes region (see Fig.~\ref{fig:haslam_zone}) permit some constraints on the flux scale factor to come from outside the unobserved region, but other choices of zone shape, size, and placement would produce different results. As a corollary, improving the calibration of the Haslam map does not strictly require full-sky observations with well-calibrated experiments. Even partial sky coverage is helpful, and can improve the constraints on the flux scale factors as long as each region has some coverage. A dedicated observation campaign to recalibrate the map could use this fact to optimise the location and survey strategy of the telescope, assuming a particular parametrisation. We emphasise that the actual spatial variation of the Haslam flux scale factor is presently unknown, however.

\begin{table}
\begin{tabular}{|c|c|c|c|c|}
\hline
{\bf Case} & {\bf FWHM {[}$\circ${]}} &  {\bf $\sigma_{\rm x}$ {[}mK{]}} & {\bf $\sigma_{\rm H}$ {[}mK{]}} & {\bf $N_{\rm sample}$} \\ \hline
{\sc STD}            & 30                           & 50                     & 800                     & 1000                \\ \hline
{\sc LXNOISE}              & --                       & 25                     & --                     & --                \\ \hline
{\sc HXNOISE}              & --                       & 175                     & --                     & --                \\ \hline

{\sc LHNOISE}              & --                       & --                     & 500                    & --               \\ \hline
{\sc CHNOISE}              & --                       & --                     & 800 [1300]                    & --               \\ \hline

\end{tabular}

\caption{The parameters used in each case in Section \ref{sec:fixed_beta} : {\sc STD} is the standard set with the default parameters; {\sc LXNOISE} is performed with lower noise rms per pixel (25~mK) for $\mathbfit{d}_x$ {\sc HXNOISE} is performed with higher noise rms per pixel (175~mK) for $\mathbfit{d}_x$; {\sc LHNOISE} is performed with lower noise (500 mK) for $\dvec_{\rm H}$; and {\sc CHNOISE} is performed with confused noise for $\dvec_{\rm H}$ where the true noise level is 1300 mK but we have ``mistakenly'' assumed it to be 800 mK. Note that there are 7 base zones, so 10 subzones means that there are 70 individual zones in total, i.e. $\gvec$ has 70 entries.}
\label{tbl:param_values}
\end{table}

We have intentionally chosen a simplistic approach to handling incomplete sky coverage here, but there are several possibilities for improving the results in regions of missing data depending on the amount of prior information and additional model structure that we are willing to incorporate. First, we adopted a simple prior on $\svec$ that assumes $\Smtrx$ is independent from pixel to pixel. As an alternative, we could have defined a prior based on the angular power spectrum, $C_\ell$, instead. This accounts for the spatially correlated nature of the temperature field, allowing neighbouring regions to inform the approximate temperature level and angular structure within missing data regions. The GCR method will then draw plausible realisations of the temperature field within the missing data regions. The power spectrum can even be included in the Gibbs scheme, allowing it to be self-consistently inferred from the data jointly with the other parameters \citep[see][]{Eriksen_2008}. Another option would have been to introduce a stronger flux scale factor prior -- essentially making a stronger assumption about the possible values of these parameters in the missing data region. We leave a more detailed exploration of missing data and inhomogeneous sensitivity to future work however.

\subsection{Varying spectral index}
\label{sec:varying_beta}

Finally, we investigate the effects of varying spectral indices on recovery of the full joint posterior distribution of the sky brightness temperature field, flux scale factors, and spectral index parameters, following the full Gibbs sampling steps described in Sections \ref{sec:gibbs} and \ref{sec:mcmc_beta}. For this study, we use the survey parameters from the {\sc STD} case.

Figure \ref{fig:skymap_spect} presents maps of the posterior mean of the sky brightness temperature $\langle \boldsymbol{s}\rangle$ for the {\sc STD} case with MCMC estimation of the six $\beta$ zones (top left) and its fractional difference with respect to the true map (top right). The difference between the uncorrected sky map (middle left) and posterior mean (middle right) with the true sky brightness temperature field are also shown, along with the posterior mean of the spectral index $\langle \boldsymbol{\beta}\rangle$ (bottom left) and its fractional difference with respect to the true values (bottom right). Although at first glance $\langle \boldsymbol{s}\rangle$ looks similar to the {\sc STD} case in Figure \ref{fig:skymap_std}, the difference from the true sky map shows that there are pixelated regions with slightly larger residuals e.g. along the North Polar Spur. These regions overlap with the beta zones, suggesting a degeneracy between the two parameters. It is also interesting to note that while $\langle \boldsymbol{\beta}\rangle$ is well within $1 \%$ of the true values, $\langle \boldsymbol{s}\rangle$ can deviate from the true value by more than 15$\%$.

% We then plot maps of the signal-to-noise ratio (SNR) of the true sky (left) and the ratio of the true sky with respect to the standard deviation of the sampled sky, $\boldsymbol{s}/\sigma_{\langle \boldsymbol{s}\rangle}$ (right) in Figure \ref{fig:snr_sigma}. The latter closely resembles the SNR map, in which the highest values are in the Galactic plane, whereas the lowest values are in regions away from the plane. 

To understand how the parameters are correlated between certain pixels, in Figure \ref{fig:corr_mtrx_zscore} we present the correlation matrix of $\langle \boldsymbol{s}\rangle, \langle \boldsymbol{g}\rangle$, and $\langle \boldsymbol{\beta}\rangle$ for a strip of 1024 consecutive pixels. Because the correlation matrix is symmetric, we will only consider the top half of the triangle. Ignoring the autocorrelation terms in the diagonal, we can see that the $\svec$ values in each pixel are generally uncorrelated with one another, as denoted by the consistently white/grey colour in the top third block of the matrix -- except for the first $\sim 140$ pixels. The $\svec$ values of these pixels are also slightly negatively correlated with the $\gvec$ and $\beta$ values. The presence of dark purple squares in the second and third blocks of the matrix indicates that some pixels are strongly positively correlated to each other. This is because they either reside in the same flux scale factor or spectral index zones. However, some $\gvec$ values are also moderately anti-correlated between different zones. Likewise, some $\beta$ zones can also be significantly anti-correlated with one another.

This complex picture illustrates the difficulty of managing high-dimensional models in the face of limited information -- in this case, due to the low angular resolution of the low-frequency experiment. Because $\beta$ and $\svec$ are correlated, this could cause significantly longer correlation lengths of the chains, reducing the effective sample size. To help address this, one could consider using analytic marginalisation, an approach taken by \cite{stompor2009MNRAS.392..216S}, or even performing a second cheap $\svec$ sampling step immediately after a $\beta$ draw. Nevertheless, we are able to successfully draw samples from the joint posterior distribution, analyse their correlations, and diagnose some of the issues that arise. This demonstrates one of the main advantages of a fully Bayesian approach -- the provision of important contextual information to help interpret the `bottom-line' results. 

Finally, in Figure \ref{fig:sed}, we present the posterior mean (squares) and true (crosses) spectral energy distributions (SEDs) of six randomly-selected pixels, one from each spectral zone (left panel), and the mean and standard deviation of their residual with respect to the true SED (right panel). The theoretical SED corresponding to the posterior mean model is plotted as solid lines using Eq.~\ref{eqn:spectra}. The residual for these pixels is $\sim 2$~K or less at the reference frequency of 408 MHz, but much larger at lower frequencies. This is to be expected given the $\sim 10-15$\% fractional differences in $\svec$ shown in Fig.~\ref{fig:skymap_spect}; the sky is much brighter at lower frequencies, and so the absolute temperature difference will be correspondingly larger.

This does point to a possible issue for attempts to build and use accurate low-frequency sky models for 21cm global signal experiments however. Even with good recovery of the high-resolution map at 408~MHz and generally good recovery of the spectral indices in each region, the large frequency range covered by the datasets, coupled with the low angular resolution of the low-frequency data, permits errors of hundreds of Kelvin below 100~MHz. While these errors should average out over the extent of the low frequency experiment's beam, they can couple with smaller-scale beam structures (e.g. in sidelobes and around nulls), which vary with frequency. This risks the introduction of spurious spectral artifacts into the data, depending on whether the sky model is used for calibration, foreground removal etc. While unable to solve this particular issue without higher-resolution data at low frequencies, our framework does at least allow these effects to be incorporated into analyses on a statistical basis.

\section{Conclusions}
\label{sec:conclusion}

Full-sky maps of radio emission play an important part in the calibration and foreground removal procedures used by 21cm arrays, CMB experiments, and many others, as well as underpinning our understanding of Galactic emission processes such as synchrotron radiation. Despite this, some of the most widely-used maps are known to harbour uncorrected systematic effects. A particular example is the 408 MHz all-sky map \citep{haslam1982A&AS...47....1H}, which was constructed from four separate surveys with three different telescopes over the course of almost two decades, starting in the 1960s. Residual striping and point source artifacts were largely corrected in \cite{remazeilles_10.1093/mnras/stv1274}, but several other effects remain poorly constrained. As one example, \cite{wilensky2024bayesian} found evidence for uncorrected flux scale errors of up to a factor of $1.6$ in three close-together pointings, based on a joint analysis with MeerKAT and LWA data. Other properties of the map, such as the beam solid angle correction and noise level, are also poorly known \citep{remazeilles_10.1093/mnras/stv1274}. In recent years, a new generation of low-angular-resolution, absolutely-calibrated radiometry experiments, targeting the sky-averaged 21cm global signal, have begun to make more accurately calibrated maps in many frequency channels across wide bands. This presents an opportunity to correct or `recalibrate' the higher-resolution diffuse maps \citep[e.g. as in][]{Monsalve_2021}, as long as a suitable multi-frequency, multi-resolution model can be defined to relate the datasets.

Several methods have been presented in the literature to model the diffuse sky in light of multiple disparate radio datasets, including accounting for systematic effects \citep[e.g.][]{Monsalve_2021, pagano_10.1093/mnras/stad3392, carter2025bayesian, 2025arXiv250621258I}. These have largely been conducted within a Bayesian statistical framework, which permits a principled treatment of model uncertainties, and regularisation of missing data through the specification of priors. The large dimensionality of the parameter space needed to fully describe the maps and data models is a considerable challenge however, and compromises such as simplified models, analytic marginalisation etc. have been made to make the problem tractable.

In this paper, we presented a Gibbs sampling scheme that permits recovery of the {\it full} joint posterior distribution of the true sky brightness temperature field $\svec$, spatially varying flux scale factors $\gvec$, and spatially varying spectral index parameters $\boldsymbol{\beta}$. In the synthetic data we used for testing, this amounts to almost 50,000 parameters, which we are able to sample in tens of seconds per iteration on a standard high-performance laptop (depending on the specific settings for the inference). In practical terms, we have shown that this is sufficient to analyse a full-sky diffuse map at a Healpix resolution of \textsc{nside}=64, with uncertain flux scale factors in 70 spatial zones, constrained by lower-resolution absolutely-calibrated data in 21 frequency channels. With further optimisation and by adding parallelism, it should be possible to scale up this proof-of-concept implementation to significantly larger datasets, and more complicated models.

After deriving the necessary mathematical results to implement the Gibbs sampler, we first tested our framework on a fiducial scenario with a fixed $\beta$ across the sky. We found that in most pixels/zones, we can constrain the posterior means $\langle \svec\rangle$ and $\langle \gvec\rangle$ to an accuracy of $\pm 10 \%$ and $\pm 2 \%$ respectively. However, certain $\gvec$ zones have larger deviations that are within $6\%$ of the true values; these zones have a substantial negative correlation with the $\svec$ parameters which likely accounts for this.

Next, we investigated the effects of different noise levels on the estimated parameters. We found that there is a negligible difference between most of the scenarios, except for when there is a lower noise in the Haslam map i.e. the {\sc LHNOISE} case, in which the constraints on $\svec$ are around $4 \%$ tighter. When there is incomplete sky coverage, we are still able to somewhat constrain the sky temperature field in the unobserved region, thanks to the flux scale factor zones allowing information to be propagated from neighbouring regions. The corrected sky brightness temperature is consistently underestimated in this region however, with a difference of $\sim5$~K compared with the true sky at 408~MHz. Other regions remain unaffected, with constraints on $\langle \boldsymbol{s}\rangle$ that are comparable to the default scenario.

In other examples, we also included the spectral index parameter $\boldsymbol{\beta}$ in the inference. We found a significant positive correlation between $\svec$ and $\boldsymbol{\beta}$, in which a difference of $< 1\%$ in the values of $\boldsymbol{\beta}$ can cause more than a $15 \%$ difference in the posterior mean of $\svec$. Nevertheless, in all cases that we considered, we are able to rectify a sky map with uncorrected flux scale factors at 408~MHz to within $\sim 5$~K or better of the true brightness temperature values.

We did not consider other foreground components such as free-free emission or point sources in our study -- the latter of which has been shown to have a non-negligible effect on 21cm global signal recovery \citep{mittal10.1093/mnras/stae2111} -- although in the future it would be possible to include multiple components within the spectral model \citep[e.g. see][]{Eriksen_2008}. For simplicity, our beam model was set to an axisymmetric Gaussian profile, and so models of the beam with sidelobes and asymmetry should also be included for increased realism. Ultimately, it would be valuable to include the beam model itself in the inference \citep{2025arXiv250620712W}.

% We have also assumed that the low frequency experiment has full sky coverage, but this is not true for global 21cm experiments like EDGES \citep{edges2018Natur.555...67B}. A better approach is to mask certain regions to mimic the coverage of actual experiments, or to include multiple datasets observed from different sites around the world to build up a full-sky reference map for calibration \citep{Ignatov:2025uzz}.
One could also consider using more frequency channels -- we assumed a relatively coarse spectral resolution of 5~MHz -- and include curvature effects and additional spatial variation of $\beta$ in the frequency spectrum model. Moreover, a more accurate prior on the flux scale factors can be used instead of the broad prior that we have chosen; this would help tighten the constraints when there is incomplete coverage of the sky. Nevertheless, the assumptions we used are appropriate and conservative for a proof-of-concept of this method, and we leave these other options to be explored in future work.

While we have only applied our method on a lower-resolution map with \textsc{nside}=64, the scheme we have presented can also be applied to a full-resolution \textsc{nside}=512 map, corresponding to $\sim 3$~million parameters. The 64-fold increase in the number of parameters would cause a significant increase in the computational time to solve for the second conditional distribution of Eq.~\ref{eq:gibbsiter}, but we expect it to be feasible. The main bottleneck is likely to be the beam convolution step, which currently benefits from assuming beam axisymmetry, so it can be performed in spherical harmonic space. This would need to be swapped for a less efficient direct convolution method for more general beam shapes. %In general, to fully utilise our Bayesian method to recalibrate the Haslam map, one would need to have low-frequency maps between 50 -- 150 MHz with a noise level of 50 mK per pixel.

For all of the results presented here, we have assumed an idealised Gaussian beam  with a FWHM of $30^\circ$. This is somewhat narrower than the beams of the compact antennas used by most 21cm global signal experiments \citep[e.g.][]{edges2018Natur.555...67B, saras2021ExA....51..193T, 2022NatAs...6..984D, 2024ApJ...961...56M}, which tend to be $60^\circ$ or larger, but is closer to the $35 - 44^\circ$ resolution of the proposed RHINO horn antenna experiment at $65-80$~MHz \citep{rhino2024arXiv241000076B}, and surpassed by the $25^\circ$ beam of the L-BASS experiment at 1.4~GHz \citep{2025RASTI...4...17Z} and $18 \times 23^\circ$ beams of the TRIS experiment at 0.6, 0.8, and 2.5~GHz \citep{2008ApJ...688...12Z}. None of the latter experiments have full-sky coverage, and they have relatively narrow bandwidths. The improved angular resolution is important for allowing the flux scale factor regions to be differentiated however, with lower-resolution experiments only permitting very coarse corrections, e.g. of a global multiplicative factor for the entire map. The real angular structure of the Haslam flux scale factor correction is unknown, but the survey region-based model we have presented here (see Fig.~\ref{fig:haslam_zone}) offers a reasonable middle ground between a coarse global correction and a detailed pixel-by-pixel correction. While the sizes and shapes of the sub-regions in Fig.~\ref{fig:haslam_zone} could be further refined, their height is best matched to an experiment with resolution between $\sim 5-10^\circ$.

Full sky coverage would require an international network of absolutely-calibrated radiometers, along similar lines to that suggested by \citet{2025arXiv250621258I}. Loosely, such a network already exists, as several existing experiments have observed from multiple sites at a range of latitudes. The necessary data are not yet publicly available however, so a real-world application of our method is not yet possible. The experiments in question are also of the very low angular resolution kind, although they do have large bandwidths; recent measurements from the GINAN experiment \citep{2025arXiv250911846M} cover $60 - 350$~MHz for example, and have been used to determine global additive and multiplicative corrections to the Global Sky Model.

\section*{Acknowledgements}

We are grateful to C.~Dickinson, M.~Irfan, D.~Watts, M.~Wilensky, and Z.~Zhang for useful discussions. This result is part of a project that has received funding from the European Research Council (ERC) under the European Union's Horizon 2020 research and innovation programme (Grant agreement No. 948764). 

%%%%%%%%%%%%%%%%%%%%%%%%%%%%%%%%%%%%%%%%%%%%%%%%%%
\section*{Data Availability}

The Gibbs sampler software presented in this paper is available from \url{https://github.com/BellaNasirudin/bayesian_skymap}.

\balance

%%%%%%%%%%%%%%%%%%%% REFERENCES %%%%%%%%%%%%%%%%%%

\bibliographystyle{mnras}
\bibliography{haslam_gains} % if your bibtex file is called example.bib

%%%%%%%%%%%%%%%%%%%%%%%%%%%%%%%%%%%%%%%%%%%%%%%%%%

% Don't change these lines
\bsp	% typesetting comment
\label{lastpage}
\end{document}